\documentclass[fleqn]{book}
\def\ang{\AA}

\def\gapprox{\lower.4ex\hbox{$\;\buildrel >\over{\scriptstyle\sim}\;$}}
\def\lapprox{\lower.4ex\hbox{$\;\buildrel <\over{\scriptstyle\sim}\;$}}

\def\ref#1{\par\noindent\hangindent1cm {#1}}

\def\captio#1{\caption{\small {#1} \normalsize}}

\usepackage{graphics}
\usepackage{graphicx}
\usepackage{epsfig}
\usepackage{times}
\usepackage{makeidx}

\begin{document}
\renewcommand{\topfraction}{0.95}
\renewcommand{\bottomfraction}{0.95}
\renewcommand{\textfraction}{0.05}
\renewcommand{\floatpagefraction}{0.95}
\renewcommand{\dbltopfraction}{0.95}
\renewcommand{\dblfloatpagefraction}{0.95}
\newcommand{\AaA}{Astron.~Astrophys.\ } 
\newcommand{\ApJ}{Astrophys.~J.\ }      
\newcommand{\GRL}{Geophys.~Res.~Lett.\ }
\newcommand{\JGR}{J.~Geophys.~Res.\ }   
\newcommand{\SP}{Solar~Phys.\ }         

\setcounter{tocdepth}{3}

 \mainmatter
 \setcounter{chapter}{12} 
\setcounter{page}{1}

\chapter{SOC Systems in Astrophysics}

\subsection*{by Markus J. Aschwanden}

\bigskip
The universe is full of nonlinear energy dissipation processes, which
occur intermittently, triggered by local instabilities,
and can be understood in terms of the {\sl self-organized criticality 
(SOC)} concept. In Table 2.1 we included
a number of cosmic processes with SOC behavior.  On the largest scale,
galaxy formation may be triggered by gravitational collapses (at least
in the top-down scenario), which form
concentrations of stars in spiral-like structures due the conservation 
of the angular momentum. Similarly, stars and planets form randomly by
local gravitational collapses of interstellar molecular clouds.
Blazars (blazing quasi-stellar objects) are active galactic nuclei
that have a special geometry with their relativistic jets pointed
towards the Earth, producing erratic bursts of synchrotron radiation
in radio and X-rays. Soft gamma repeaters are strongly magnetized
neutron stars that produce crust quakes (in analogy to earthquakes)
caused by magnetic stresses and star crust fractures. Similarly,
pulsars emit giant pulses of radio and hard X-ray bursts during
time glitches of their otherwise very periodic pulsar signal.
Blackhole objects are believed to emit erratic pulses by magnetic
instabilities created in the accretion disk due to rotational shear
motion. Cosmic ray particles are the result of a long-lasting series
of particle acceleration processes accumulated inside and outside of 
our galaxy, which is manifested in a powerlaw-like energy spectrum
extending over more than 10 orders of magnitude. Solar and stellar
flares are produced by magnetic reconnection processes, which are
observed as impulsive bursts in many wavelengths. Also phenomena
in our solar system exhibit powerlaw-like size distributions, such
as Saturn ring particles, asteroids, or lunar craters, which are
believed to be generated by collisional fragmentation processes or
their consequences (in form of meteroid impacts). The magnetosphere
of planets spawns magnetic reconnection processes also, giving rise
to substorms and auroras. 

While these astrophysical processes have been interpreted in terms of
the self-organized criticality concept (for a comprehensive overview
see Aschwanden 2011a), quantitative theoretical modeling of astrophysical
SOC phenomena is still largely unexplored. In the following Section 13.1
we outline a general theory approach to SOC phenomena, which consists 
of a universal (physics-free) mathematical/statistical aspect, as well as a 
physical aspect that is unique to each astrophysical SOC phenomenon or 
observed wavelength range. In the subsequent Section 13.2 we discuss 
then the astrophysical observations and compare the observed size 
distributions with the theoretically predicted ones.  

\section{Theory}

A system with nonlinear energy dissipation governed by self-organized 
criticality (SOC) is usually modeled by means of cellular automaton (CA)
simulations (BTW model; Bak et al.~1987, 1988). A theoretical definition 
of a SOC system thus can be derived from the mathematical rules of a CA 
algorithm, which includes: (1) A S-dimensional rectangular lattice grid, 
(2) a place-holder for a physical quantity $z_{i,j,k}$ associated with 
each cellular node $x_{i,j,k}$, (3) a definition of a critical threshold 
$z_{crit}$, (4) a random input $\Delta z_{i,j,k}$ in space and time; 
(5) a mathematical re-distribution rule that is applied when a local 
physical quantity exceeds the critical threshold value which adjusts 
the state of the next-neighbor cells, and (6) iterative time steps 
to update the system state $z_{i,j,k}(t)$ as a function of time $t$.
Although this definition is sufficient to set up a numerical simulation
that mimics the dynamical behavior of a SOC system, it does not quantify
the resulting powerlaw-like size distributions in an explicit way, nor
does is include any physical scaling law that is involved in the 
relationship between statistical SOC parameters and astrophysical 
observables. A quantitative SOC theory should be generalized in such 
a way that it encompasses both the mathematical/statistical aspects of
a SOC system, as well as the physical scaling laws between observables
and statistical SOC parameters. In the following we generalize the
fractal-diffusive SOC model (FD-SOC), described in Aschwanden (2012a)
and outlined in Section 2.2.2, which includes three essential parts:
two universal statistical aspects, i.e., (i) the {\sl scale-free
probability theorem}, (ii) the {\sl fractal-diffusive spatio-temporal 
relationship}, and a physical aspect, i.e., (iii) physical scaling laws 
between geometric SOC parameters and astrophysical observables, which
may be different for each observed SOC phenomenon and each observed 
wavelength in astrophysical data. Some basic examples of physical
scaling laws are derived for fragmentation processes, for thermal 
emission of astrophysical plasmas, and for astrophysical particle 
acceleration mechanisms.

\begin{figure}
\centerline{\includegraphics[width=0.6\textwidth]{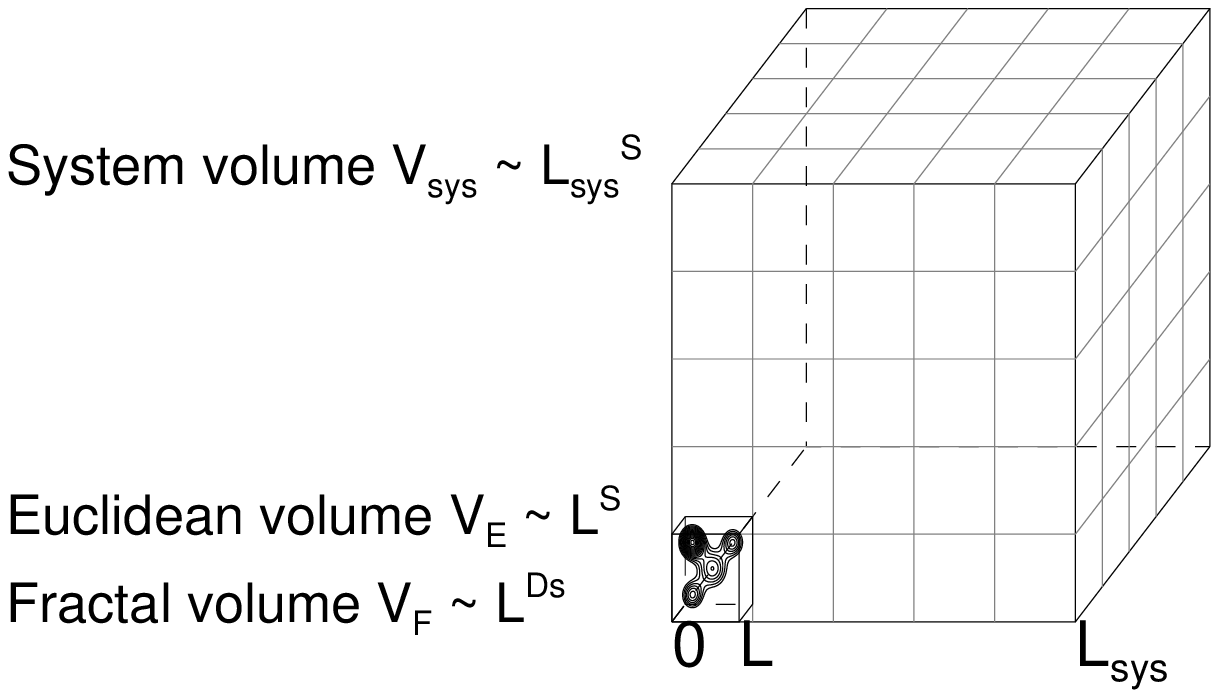}}
\captio{The geometric relationships between the 
Euclidean avalanche volume $V_E \propto L^S$, the fractal 
avalanche volume $V \propto L^{D_s}$, and the system volume
$V_{sys} \propto L_{sys}^S$ is visualized in 3D space $(S=3)$,
as a function of the avalanche length scale $L$ and system size
$L_{sys}$. Note that the probability for an avalanche with size $L$
scales with the ratio of the system volume $V_{sys}$ to the
Euclidean avalanche volume $V_E$.}
\end{figure}

\subsection{The Scale-Free Probability Theorem} 

Powerlaw-like size distributions are an omnipresent manifestation
of SOC phenomena, a property that is also called a ``scale-free''
parameter distribution, because no preferred scale is singled out
by the process. Of course, the scale-free parameter range, over which
a size distribution exhibits a powerlaw function, is always limited by
instrumental sensitivity or a detection threshold at the lower end, 
and by the finite length of the time duration over which a SOC system 
is observed and sampled, at the upper end. 
Bak et al.~(1987, 1988) associated the scale-free behavior 
with the fundamental property of 1/f-noise that is omnipresent in
many physical systems, giving rise to a power spectrum of $P(\nu)
\propto \nu^{-1}$. 

However, here we give a more elementary explanation
for the powerlaw behavior of SOC size distributions, namely in terms
of the statistical probability for scale-free avalanche size distributions. 
A key property of SOC avalanches is that random disturbances can 
produce both small-scale as well as unpredictable large-scale avalanches 
of any size, within
the limitations of a finite system size $L_{sys}$ at the upper end,
and some ``atomic'' graininess $\Delta L$ at the lower end 
(i.e., a sand grain in sand 
avalanches, or the spatial pixel size $\Delta L$ in a computer lattice
grid). The statistical probability distribution 
$N(L)$ for avalanches with size $L$ can be calculated from the statistical
probability. If no particular size is preferred in a scale-free process,
the number $N(L)$ of possible avalanches in a S-dimensional system with 
a volume $V_{sys} = L_{sys}^S$ is simply the system volume $V_{sys}$
divided by the Euclidean volume $V_E=L^S$ of a single avalanche with length 
scale $L$ (Fig.~13.1),
$$
	N(L)  \propto \left( {L_0 \over L} \right)^S \propto L^{-S} \ , 
	\eqno(13.1)
$$
In a slowly-driven SOC system, none or only one avalanche happens at one 
particular time, but the relative probability that an avalanche can happen
still scales with the reciprocal volume. This is our first basic assumption
of our generalized SOC model, which we call the {\sl scale-free probability
theorem}. This theorem itself predicts a powerlaw function for the
basic size distribution of spatial scales, and is distinctly different from
the gaussian distribution function that results from binomial probabilities.
For 3-dimensional SOC phenomena ($S=3$), thus we expect a size distribution
or differential occurrence frequency distribution of $N(L) \propto L^{-3}$,
or a cumulative occurrence frequency distribution of $N(>L) \propto L^{-2}$.

We can now derive the expected size distributions for related geometric
parameters, such as for the Euclidean avalanche area $A_E$ or volume $V_E$. 
If we simply define the Euclidean area $A_E$ in terms of the squared length 
scale, i.e., $A_E \propto L^2$ or $L \propto A_E^{1/2}$, which has the 
derivative $dL/dA_E \propto A_E^{-1/2}$, we obtain for the area size 
distribution $N(A_E)$, using $N(L) \propto L^{-S}$ (Eq.~13.1), 
$$
	N(A_E) dA_E \propto N[L(A_E)] \left| {dL \over dA_E} \right| dA_E 
	\propto A_E^{-(1+S)/2} dA_E \ ,
	\eqno(13.2)
$$
yielding $N(A_E) \propto A_E^{-2}$ for 3D phenomena ($S=3)$.
Similarly we define the Euclidean volume, i.e., $V_E \propto L^3$ or
$L \propto V_E^{1/S}$, which has the derivative 
$dL/dV_E \propto V^{1/S-1}$, yielding a volume size distribution $N(V_E)$,  
$$
	N(V_E) dV_E \propto N[L(V_E)] \left| {dL \over dV_E} \right| dV_E 
	\propto V_E^{-(2-1/S)} dV_E \ ,
	\eqno(13.3)
$$
yielding $N(V_E) \propto V_E^{-5/3}$ for 3D phenomena ($S=3)$.

In the case that avalanche volumes are fractal, such as characterized with
a Hausdorff dimension $D_S$ in Euclidean space with dimension $S$, the
fractal volume $V$ scales as,
$$
	V \propto L^{D_S} \ ,
	\eqno(13.4)
$$
which yields $L \propto V^{1/D_S}$ and the derivative $dL/dV
\propto V^{(1/D_S -1 )}$, and
thus the size distribution, 
$$
	N(V) dV \propto N[L(V)] \left| {dL \over dV} \right| dV 
	            \propto V^{-[1+(S-1)/D_S]} \ .
	\eqno(13.5)
$$
The Euclidean limit of non-fractal avalanches would yield for $S=3$
and $D_S=S=3$ the same powerlaw index $\alpha_V=1+(S-1)/D_S=5/3$ as
derived for $V_E$ in Eq.~(13.3). For fractal avalanches, with
$D_S \approx (1+S)/2=2.0$ for $S=3$, we obtain a slightly steeper
powerlaw distribution, $N(V) \propto V^{-2.0}$, than for Euclidean
avalanches, i.e., $N(V_E) \propto V_E^{-5/3}$. 

A similar effect occurs for fractal avalanche areas $A$. If we
assume an fractal structure with Hausdorff dimension $D_2$ in 2D space
$$
	A \propto L^{D_2} \ ,
	\eqno(13.6)
$$
which yields $L \propto A^{1/D_2}$ and the derivative
$dL/dA=A^{1/D_S-1}$, and thus a size distribution of 
$$
	N(A) dA \propto N[L(A)] \left| {dL \over dA} \right| dA 
	            \propto A^{-[1+(S-1)/D_2)]} \ .
	\eqno(13.7)
$$
For fractal avalanches, with $D_2 \approx (1+S)/2=1.5$ for the
2-D fractal dimension, 
we obtain a powerlaw distribution $N(A) \propto A^{-7/3}$, 
which is slightly steeper than for Euclidean avalanche areas, i.e., 
$N(A_E) \propto A_E^{-2.0}$. 

In some astrophysical observations, such as in solar and planetary physics,
the areas $A$ of SOC phenomena can be measured, while spatially integrated
emission (or spatially unresolved emission) in (optically-thin) soft X-ray or 
extreme-ultraviolet wavelengths is often roughly proportional to the 
fractal volume 
$V$ of the emitting source, and thus the size distributions Eqs.~(13.5) and
(13.7) can be used to test our scale-free probability theorem (Eq.~13.1). 

If the scale-free probability theorem (Eq.~13.1) is correct, the derived size
distributions for spatial scales $N(L)$, avalanche areas $N(A)$, and avalanche 
volumes $N(V)$, should be universally valid for
SOC phenomena, without any physical scaling laws. They should be
equally valid for earthquakes or solar flares, regardless of the
physical mechanism that is involved in the nonlinear energy dissipation
process of a SOC event. In Section 13.2 we will present some astrophysical
measurements of size distributions of such geometric parameters ($L, A$) 
which can corroborate our assumption of the
scale-free probability theorem. Volume parameters ($V$)
can usually not directly be measured for astrophysical objects,
except by means of stereoscopy or tomography of nearby objects.

\subsection{The Fractal-Diffusive Spatio-Temporal Relationship} 

After we have established a framework for the statistics of spatial 
or geometric parameters of SOC avalanche events, we turn now to temporal
parameters, which can be defined by a spatio-temporal relationship. The 
temporal evolution of SOC avalanches is governed by the complexity
of next-neighbor interactions above some threshold value, which
has an erratically fluctuating time characteristics according to
cellular automaton simulations. However, the mean radius $r(t)$ of
an evolving SOC avalanche was found to closely mimic a time dependence
of $r(t) \propto t^{1/2}$, which can be associated with a classical
random-walk or diffusion process (Aschwanden 2012a). Measurements of
the spatial evolution of solar flares, which are considered to be an 
established SOC phenomenon, revealed a similar evolution, but tend
to be sub-diffusive for the analyzed dataset (Aschwanden 2012b).
Moreover, the instantaneous avalanche area was found to have a
fractal structure, while the time-integrated avalanche area is nearly
space-filling and thus can be described with an Euclidean area or
volume. These two properties of fractal geometry and diffusive
evolution have been combined in the {\sl fractal-diffusive SOC avalanche
model (FD-SOC)} (Aschwanden 2012a). We generalize this concept now
also for anomalous diffusion,
$$
	r(t) \propto \kappa \ t^{\beta/2} \ ,
	\eqno(13.8)
$$
where $\kappa$ is the diffusion coefficient and the diffusive index
$\beta$ combines classical diffusion ($\beta=1$), as well as anomlous
diffusion ($\beta \ne 1$). Anomalous diffusion processes include 
both sub-diffusion ($\beta=0...1$), as well as super-diffusion 
($\beta=1...2$), also called hyper-diffusion or L\'evy flights,
$$
        r(t) \propto t^{\beta/2} \qquad
        \left\{
        \begin{array}{ll}
                \beta < 1 & \mbox{(sub-diffusion)} \\
                \beta = 1 & \mbox{(classical diffusion)} \\
                \beta > 1 & \mbox{(super-diffusion or L\'evy flights)}
        \end{array}
        \right.
	\eqno(13.9)
$$
We show the generic time evolution of a sub-diffusion process with
$\beta=1/2$ and a super-diffusion process with $\beta=3/2$ in Fig.~13.2.
Anomalous diffusion implies more complex properties of the diffusive
medium than a homogeneous structure, which may include an
inhomogeneous fluid or fractal properties of the diffusive medium.

The spatio-temporal evolution of an instability generally starts
with an exponential growth phase (which we may call the acceleration
phase), followed by a saturation or quenching phase (which we may call
deceleration phase). In the logistic growth model (Section 2.2.1),
the deceleration phase saturates asymptotically at a fixed value,
while diffusive models do not converge but slow down progressively
with time (Fig.~13.2).

\begin{figure}
\centerline{\includegraphics[width=0.9\textwidth]{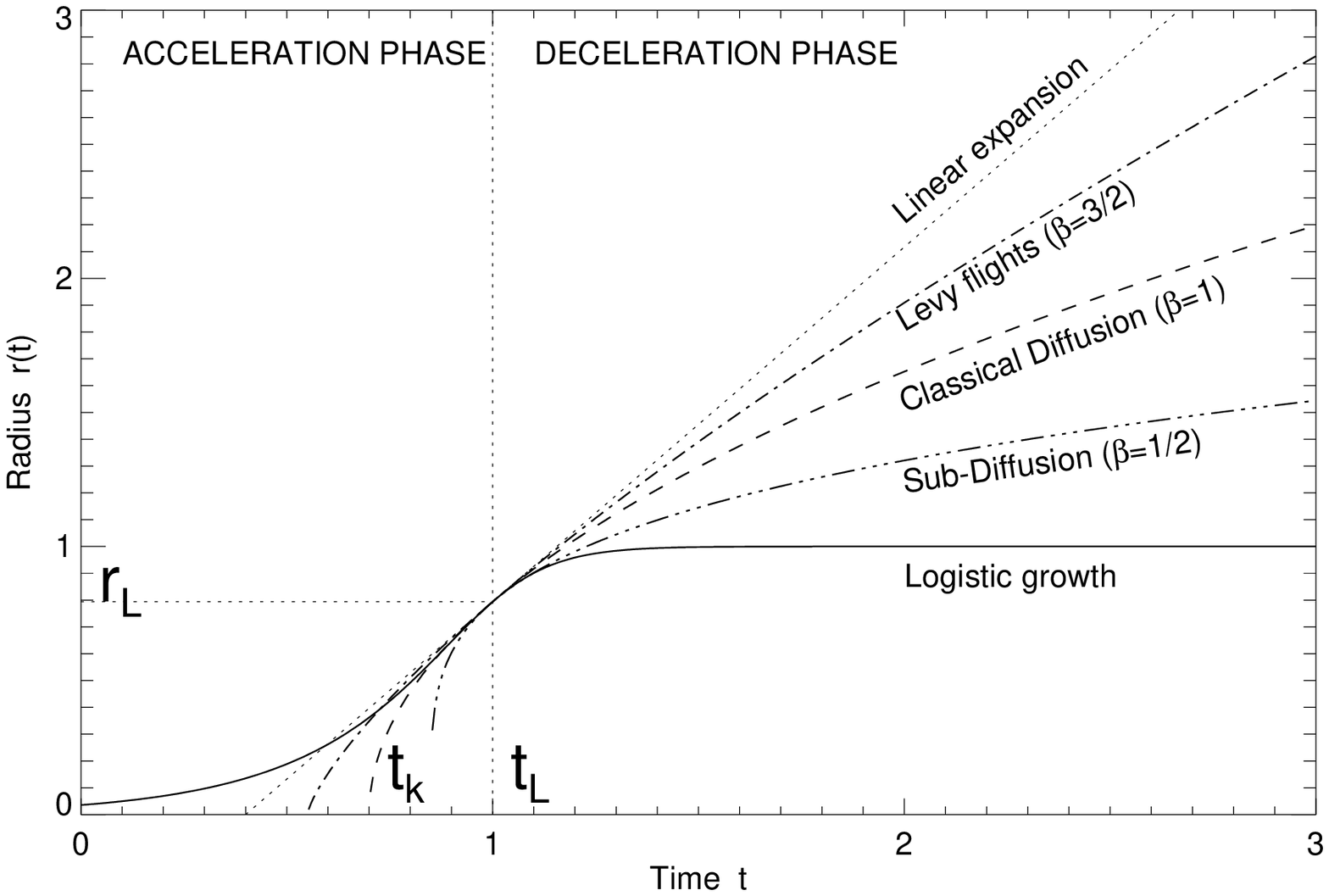}}
\captio{Comparison of spatio-temporal evolution models:
Logistic growth with parameters $t_L=1.0, r_\infty=1.0, \tau_G=0.1$,
sub-diffusion ($\beta=1/2$), classical diffusion ($\beta=1$),
L\'evy flights or super-diffusion ($\beta=3/2$), and linear expansion
($r \propto t$).  All curves intersect at $t=t_L$
and have the same speed $v=(dr/dt)$ at the intersection point at
time $t=t_L$. (Aschwanden 2012b).}
\end{figure}

The diffusive scaling (Eq.~13.9) implies then also a statistical
correlation between spatial $L$ and temporal scales $T$,
$$
	L \propto T^{\beta/2} \ ,
	\eqno(13.10)
$$
where $T$ is the time duration of a SOC avalanche. From the
scale-free probability theorem $N(L) \propto L^{-S}$ (Eq.~13.1) we can 
then directly compute the expected occurrence frequency distribution $N(T)$ 
for time durations $T$,
$$
	N(T) dT \propto N[L(T)] \left|{dL \over dT} \right| dT 
	            \propto T^{-[1+(S-1) \beta / 2]} \ dT \ .
	\eqno(13.11)
$$
For instance, for 3D SOC phenomena ($S=3$) we expect a powerlaw 
distribution $N(T) \propto T^{-(1+\beta)}$, which amounts to 
$N(T) \propto T^{-2.0}$ for classical diffusion, 
$N(T) \propto T^{-1.5}$ for a sub-diffusion case ($\beta=0.5$), or
$N(T) \propto T^{-2.5}$ for a super-diffusion case ($\beta=1.5$). 
The exponentially growing phase is neglected in this derivation,
which implies a slight underestimate of the number of short time scales.
Time scales can also directly be measured in most SOC phenomena,
and thus provide an immediat test of the fractal-diffusive assumption
made here, regardless of the physical process that is involved
in the observed signal of SOC avalanches. 

\subsection{Size Distributions of Astrophysical Observables}

The previous theory on geometric $(L, A, V)$ and temporal $(T)$ parameters
should be universally valid for the statistics of SOC phenomena, and thus
constitutes a purely ``physics-free'' mathematical or statistical property 
of SOC systems. All other observables of SOC events, however, are related
to a physical (nonlinear) energy dissipation process, which needs to be modeled
in terms of a correlation or scaling law with respect to the physics-free
spatio-temporal SOC parameters. Say, if we observe a physical SOC variable
$x$ that has a correlation or powerlaw scaling law of $x \propto L^\gamma$
with the geometric SOC parameter $L$, we can infer the expected size
distribution $N(x) dx$ by substituting the scaling law. 

What is most common in astrophysical observations is the flux $F$ or
intensity that is observed in some wavlength range $\lambda$ (with physical
units of energy per time), originating
from a source with unknown volume $V$. The flux $F(t)$ can exhibit
strong fluctuations during an energy dissipation event, but we can 
characterize the time profile of the event with a peak flux $P$, or
with the time-integrated flux, also called fluence (with physical untis of
energy), which we may denote with $E$. For optically-thin
emission observed in soft X-ray or EUV emission, the emissivity or flux
is approximately proportional to the source volume $V$, so it is most
useful to quantify a scaling law of the flux $F$ with the 3D volume $V$,
which we characterize with a powerlaw index $\gamma$,
$$
	F \propto V^{\gamma} \ .
	\eqno(13.12)
$$
From the size distribution of the fractal volume 
$N(V) \propto V^{-[1+(S-1)/D_S]}$ (Eq.~13.5) and the scaling law
$V \propto F^{1/\gamma}$ (Eq.~13.12) and its derivative 
$dV/dF \propto F^{1/\gamma-1}$ we can then derive the size distribution
$N(F)$ of fluxes $F$,
$$
	N(F) dF \propto N[V(F)] \left| {dV \over dF} \right| dF 
	            \propto F^{-[1+(S-1) / (\gamma D_S)]} dF \ .
	\eqno(13.13)
$$
which has a typical powerlaw index of $\alpha_F \approx 2.0$ (for
$S=3$, $D_S\approx (1+S) /2 \approx 2.0$, and $\gamma \approx 1$).
For peak fluxes $P$ we have the same distribution, except for the
fractal dimension having its maximum Euclidean value $D_S \approx S$,
which yields,
$$
	N(P) dP \propto N[V(P)] \left| {dV \over dP} \right| dP 
	            \propto P^{-[1+(S-1) / \gamma S]} dF \ .
	\eqno(13.14)
$$
which has a typical value of $\alpha_P \approx 5/3 \approx 1.67$. 
Finally, the total flux or fluence $E = \int F(t) dt \approx F T$,
is found to have a size distribution of,
$$
	N(E) dE \propto N[V(E)] \left| {dV \over dE} \right| dE 
	            \propto E^{-[1+(S-1) / (\gamma D_S + 2/\beta)]} dE \ .
	\eqno(13.15)
$$
which has a typical value of $\alpha_E \approx 3/2 = 1.5$,
for $S=3$, $D_S \approx (1+S)/2 \approx 2.0$, $\gamma \approx 1$, and
$\beta \approx 1$.  

In summary, if we denote the occurrence frequency distributions
$N(x)$ of a parameter $x$ with a powerlaw distribution with power index
$\alpha_x$,
$$
        N(x) dx \propto x^{-\alpha_x} \ dx \ ,
        \eqno(13.16)
$$
we have the following powerlaw coefficients $\alpha_x$ for the parameters
$x=L, A, V, T, F, P$, and $E$,
$$
        \begin{array}{ll}
        \alpha_L     &=  S                            \\
        \alpha_A     &=  1+(S-1) / D_2                \\
        \alpha_V     &=  1+(S-1) / D_S                \\
        \alpha_T     &=  1+(S-1) \beta/2              \\
        \alpha_F     &=  1+(S-1)/(\gamma D_S)         \\
        \alpha_P     &=  1+(S-1)/(\gamma S)           \\
        \alpha_E     &=  1+(S-1)/(\gamma D_S +2/\beta) \\
        \end{array} \ .
        \eqno(13.17)
$$
Thus, the various powerlaw indices depend on four
fundamental parameters: the Euclidean dimension $S$ of the SOC system,
the fractal dimension $D_S$ of SOC avalanches, the diffusion index
$\beta$ of the SOC avalanche evolution, and the scaling law index
$\gamma$ between the observed flux and the SOC avalanche volume.
Note, that the powerlaw slopes of the geometric ($L, A, V$) and 
flux parameters ($F, P$) do not depend on the diffusion index $\beta$, 
and thus are identical for classical or anomalous diffusion.
Only the powerlaw slopes of the length scale, time scale, 
and peak flux ($L,T,P$)
do not depend on the fractal dimension. All flux-related parameters
($F, P, E$) depend on a physical scaling law ($\gamma$), which may be
different for every observed wavelength range. 

In the following we generally assume 3D SOC phenomena ($S=3$),
for which the fractal dimension can be estimated by the mean value
between the minimum and maximum dimension where SOC avalanches can
propagate coherently via next-neighbor interactions (which limits 
the minimum fractal dimension to $D_{S,min} \approx 1.0$
and maximum fractal dimension to $D_{max} = S$),
$$
	D_S \approx {D_{S,min} + D_{s,max} \over 2 } = {(1 + S) \over 2} \ ,
	\eqno(13.18)
$$
which yields $D_3 \approx 2.0$ for $S=3$ and simplifies the powerlaw
indices to 
$$
        \begin{array}{ll}
        \alpha_L     &=  2                                 \\
        \alpha_A     &=  7/3                               \\
        \alpha_V     &=  2                                 \\
        \alpha_T     &=  1+\beta               \approx 2   \\
        \alpha_F     &=  1+1/\gamma            \approx 2   \\
        \alpha_P     &=  1+2/(3 \gamma)        \approx 5/3 \\
        \alpha_E     &=  1+1/(\gamma+1/\beta)  \approx 3/2 \\
        \end{array} \ .
        \eqno(13.19)
$$
In astrophysics, the distributions of geometric parameters ($L, A, V$)
can only be determined from imaging observations with sufficient spatial
resolution (in magnetospheric, heliospheric, and solar physics), while
the distributions of all other parameters ($T, F, P$, and $E$) can be 
measured from any non-imaging observations, such as from point-like 
stellar objects. 

\subsection{Scaling Laws for Thermal Emission of Astrophysical Plasmas} 

Solar flares and stellar flares are observed in soft X-ray and 
extreme-ultraviolet (EUV) wavelengths, where the observed intensity
is measured in a particular wavelength range $\lambda$ given by the 
instrumental filter response function. Soft X-ray and EUV emission
is produced by photons via the free-free bremsstrahlung process,
free-bound transitions, or radiative recombination. In strong magnetic
fields, cyclotron and gyrosynchrotron emission is also produced at radio
wavelengths. Soft X-ray and EUV emission occur usually in the optically
thin regime, and thus the total emission measure $EM$, which is proportional 
to the observed intensity in a given wavelength $\lambda$, is proportional 
to the volume $V$ of the emitting source,
$$
	F_{\lambda} \propto EM = \int n_e^2({\bf x}) \ dV 
		\propto\ \langle n_e^2 \rangle \ V \ . \\
	\eqno(13.20)
$$
Thus, if the electron density $n_e$ in a source would be constant,
or the same among different flare events, we would have just the
simple relation $F_{\lambda} \propto V^{\gamma}$ (Eq.~13.12)
with the scaling law index $\gamma=1$, which is an approximation 
that is often made. In fact, this is quite a reasonable approximation for
measurements with a narrow-band temperature filter, which is sensitive 
to a particular electron temperature $T_e$, and thus probes also
a particular range of electron densities $n_e$ and plasma pressure 
$p = 2 n_e k_B T_e$ that depend on this temperature $T_e$,
whatever the scaling law between electron temperature $T_e$ and electron
density $n_e$ is.

The proportionality constant between the flux intensity $F_{\lambda}$ and 
emission measure $EM$ is dependent on the wavelength range $\lambda$, because
each wavelength filter is centered around a different temperature range $T_e$ 
that corresponds to the line formation temperature in the observed
wavelengths $\lambda$. Physical scaling laws have been derived to quantify
the relationship between electron temperature $T_e$, electron density
$n_e$, and the spatial length scale $L_{loop}$ of coronal loops, e.g., 
by assuming a balance between the heating rate, conductive, and 
radiative loss rate (i.e., the so-called RTV law; Rosner, Tucker, and 
Vaiana 1978), being (in cgs-units), 
$$
	T_e \approx 1400 \ (p L_{loop})^{1/3} \ .
	\eqno(13.21)
$$
which we can express in terms of the electron density $n_e$, using the 
definition of the ideal gas law, $p = 2 n_e k_B T_e$, 
$$
	n_e \approx 1.3 \times 10^9
	\left( {T_e \over 1\ {\rm MK}} \right)^2
	\left({ L_{loop} \over 10^9 \ {\rm cm}} \right)^{-1} 
	\quad [{\rm cm}^{-3}] \ .
	\eqno(13.22)
$$
Thus, if there is no particular correlation between the loop length $L_{loop}$
of the densest flare loops (with the highest emission measure) and the
volume $V$ of the active region, the density is only a function of the
electron temperature, $n_e \propto T_e^2$. Consequently, if a narrowband
temperature filter is used in a soft X-ray or EUV wavelength range,
sensitive to a peak temperature $T_\lambda$, the corresponding electron
density is given in a narrow range also, $n_e \propto T_\lambda^2$, and
thus the flux is essentially proportional to the flare volume
(according to Eq.~13.20),
$F_{\lambda} \propto V^\gamma$, with a scaling index of $\gamma \approx 1$.
On the other hand, if a different powerlaw index $\gamma \neq 1$ is measured, 
such an observation would reveal a systematic scaling of some parameters
($T_e, n_e, L_{loop}$) of the densest flare loops with the size $L$ of 
the active region.

Another important quantity we want to calculate is the size distribution 
of thermal energies $E_{th}$, for which we expect a scaling law of (using
Eq.~13.22),
$$
	E_{th} = 3 n_e k_B T_e V \propto T_e^3 V / L_{loop}\ ,
	\eqno(13.23)
$$
where the most dominant value of the electron temperature $T_e$ is given
by the peak of the {\sl differential emission measure distribution (DEM)}.
Observationally, it was found that the DEM peak temperature of a flare
scales approximately with the size $L$ of a flare, i.e., $T_e \propto L$ 
(Aschwanden 1999), which yields with 
$V \propto L^{D_S}$ (assuming that $L_{loop}$ with the highest
emission measure is uncorrelated with the active region size $L$),
$$
	E_{th} \propto L^{3+D_S} \ ,
	\eqno(13.24)
$$
or $E_{th} \propto L^5$ for $D_S\approx 2.0$ and $S=3$. The size
distribution for thermal energies is then expected to be,
$$
	N(E_{th}) dE_{th} = N(L[E_{th}]) \left| {dL \over dE_{th}} \right| 
	dE_{th} =E_{th}^{-[1+(S-1)/(3+D_S)]} dE_{th}
	\eqno(13.25)
$$
which yields $N(E_{th}) \propto E_{th}^{-1.4}$ for $S=3$ and $D_S\approx 2.0$.
However, we have to be aware that this estimate of the thermal energy
contained in a flare requires multi-thermal measurements to derive the
peak DEM temperature and cannot be obtained from a single narrowband filter
measurement. 
Note that the size distribution of thermal energies with powerlaw index
$\alpha_{E_{th}}=1.4$ is very similar to the total energy in photons in
any wavelength range, i.e., fluence, $\alpha_E=1.5$ (Eq.~13.19). 

The foregoing model for the thermal energy requires a scaling law
between the flare size $L$ and its statistical temperature $T_e$.
In practice, however, thermal energies were often estimated in a limited
temperature range from the filter ratio of two narrowband filters. 
Such filter ratio measurements are sensitive to a particular 
temperature $T_e$ and electron density $n_e$ (Eq.~13.22), which are
then essentially constants in the expression for the thermal energy,
and thus the thermal energy is mainly proportional to the volume,
$$
	E_{th,V} \propto V \ ,
	\eqno(13.26)
$$
so that the size distribution of thermal energies $V$ is identical to the
size distribution of volumes $V$, which has a powerlaw slope of
$\alpha_{th,V}=\alpha_V=1+(S-1)/D_3$, yielding values in the range
of $\alpha_{th,V}=1.67-2.0$, depending on fractal ($D_3 \approx 
(1+S)/2 = 2$ for $S=3$) or Euclidean ($D_3=S=3$) volume measurements.

In some studies, the volume is approximated with a ``pill-box'' geometry,
i.e., the product of the measured flare area $A$ with a constant height $h$
along the line-of-sight, $V = A h$, which makes the volume proportional
to the area, and consequently the thermal energy is mainly proportional
to the flare area $A$,  
$$
	E_{th,A} \propto A \ ,
	\eqno(13.27)
$$
leading to a size distribution of $E_{th,A}$ that is identical with that
of the area $A$, which has the powerlaw slope $\alpha_{th,A} = \alpha_A
=1+(S-1)/D_2$, yielding values in the range 
of $\alpha_{th,A}=2.0-2.3$, depending on fractal 
($D_2 \approx (1+S)/2 = 1.5$) or Euclidean ($D_2=2$) area measurements
in $S=3$ space.
Thus, accurate measurements of these size distributions provide a
sensitive tool to diagnose the underlying physical scaling laws
and model assumptions. 

\subsection{Scaling Laws for Astrophysical Acceleration Mechanisms} 

Let us consider some simple examples of particle acceleration processes in
astrophysical plasmas, such as solar flares, stellar flares, or cosmic rays.
The simplest particle acceleration process is a coherent direct current (DC)
electric field, which can be characterized by an acceleration constant $a$
over a system length $L$. Newtonian (non-relativistic) mechanics 
predicts for an electron with mass $m_e$ a velocity $v=a t$ after a 
distance $L=(1/2)at^2$, which corresponds to a kinetic energy $E_L$ of
(where the subscript $L$ refers to the length scale of the accelerator),
$$ 
	E_L = {1\over 2} m_e v^2 = m_e a L \ ,
	\eqno(13.28)
$$
which implies a linear energy increase with system length,
$E_L \propto L$. Thus, the size distribution of energies $E_L$
is identical with that of length scales $L$, which we obtain from the
scale-free probability theorem $N(L) \propto L^{-S}$ (Eq.~13.1),
$$
	N(E_L) dE_L = N(L[E_L]) \left| {dL \over dE_L} \right| dE_L
			= E_L^{-S} dE_L \ ,
	\eqno(13.29)
$$
yielding an energy spectrum $N(E_L) \approx E_L^{-3}$
for a 3D Euclidean volume. This simplest case is given by coherent
acceleration thoughout the entire source volume.

Another approach could be made by assuming that the magnetic energy
density $dE_B/dV = B^2/8\pi$ per volume element converted into nonthermal 
particle energy is uncorrelated with the system size $L$, in which case 
the total converted magnetic energy just scales with the Euclidean 
volume $V \propto L^S$ of the energy dissipation process, 
$E_B \propto V_E \propto L^S$. The corresponding size distribution 
of magnetic energy $E_B$ is then expected to scale as,
$$
	N(E_{B}) dE_{B} = N(L[E_{B}) \left| {dL \over dE_{B}} \right| dE_{B}
			= E_{B}^{-(2-1/S)} dE_{B} \ ,
	\eqno(13.30)
$$
which translates into $N(E_B) \propto E_B^{-5/3}$ for a 3D 
Euclidean volume (S=3). Interestingly, this value is also identical
with the size distribution of the peak flux $P$. Alternative scaling
laws using the Alfv\'enic crossing time through the flare volume
have also been considered (e.g, Shibata and Yokoyama 1999, 2002;
Nishizuka et al.~2008).

Thus, the two acceleration models yield powerlaw size distributions
in the range of $\alpha_{E_B}=1.67$ to $\alpha_{E_L}=3.0$. 
The measurement of size distributions of nonthermal energies thus
can yield valuable diagnostics about the physical nature of the
underlying particle acceleration process. 

A summary of all theoretically derived powerlaw indices expected
in astrophysical systems is compiled in Table 1. 

\begin{table}[h]
\begin{center}
\normalsize
\captio{Summary of powerlaw indices predicted in astrophysical systems,
as a function of the dimensionality $S$, the fractal dimension $D_S$,
the diffusion power index $\beta$, and the energy-volume scaling index 
$\gamma$.}
\medskip
\begin{tabular}{|l|l|l|}
\hline
Parameter		& Powerlaw index	& Powerlaw index for	  \\
			& (general expression)	& $S=3$, $D_3=2$, $D_2=3/2$ \\
			&                       & $\beta=1$, $\gamma=1$ \\
\hline
\hline
Length scale $L$	& $\alpha_L=S$ 		& $\alpha_L=3$	      \\
Area $A$                & $\alpha_A=1+(S-1)/D_2$ & $\alpha_A=7/3$     \\
Volume $V$              & $\alpha_V=1+(S-1)/D_3$ & $\alpha_V=2$       \\  
Time duration $T$       & $\alpha_T=1+\beta$    & $\alpha_T=2$        \\
Flux $F$		& $\alpha_F=1+1/\gamma$   & $\alpha_F=2$        \\
Peak flux $P$		& $\alpha_P=1+2/(3 \gamma)$ & $\alpha_P=5/3$ \\
Fluence $E$		& $\alpha_E=1+1/(\gamma+1/\beta)$ & $\alpha_E=3/2$\\
Emission measure $EM_\lambda$ & $\alpha_{EM_\lambda}=\alpha_V$ 
	& $\alpha_{EM_\lambda}=2$ \\
Thermal energy $E_{th}$ & $\alpha_{E_{th}}=1+(S-1)/(3+D_S)$ &
			  $\alpha_{E_{th}}=7/5$ \\
Thermal energy $E_{th,A}$ & $\alpha_{E_{th,A}}=\alpha_A$ &
			  $\alpha_{E_{th}}=7/3$ \\
Thermal energy $E_{th,V}$ & $\alpha_{E_{th,V}}=\alpha_V$ &
			  $\alpha_{E_{th}}=2$ \\
Linear energy $E_L$     & $\alpha_{E_L}=S$
			& $\alpha_{E_L}=3$ \\
Magnetic energy $E_B$   & $\alpha_{E_B}=(2-1/S)$
			& $\alpha_{E_B}=5/3$ \\
\hline
\end{tabular}
\end{center}
\end{table}

\section{Observations}

We discuss now a number of astrophysical observations with regard to
their observed size distributions, which we compare with the foregoing
theoretical predictions. A more detailed review of these measurements
is given in Aschwanden (2011; chapters 7 and 8). 

\begin{figure}
\centerline{\includegraphics[width=1.0\textwidth]{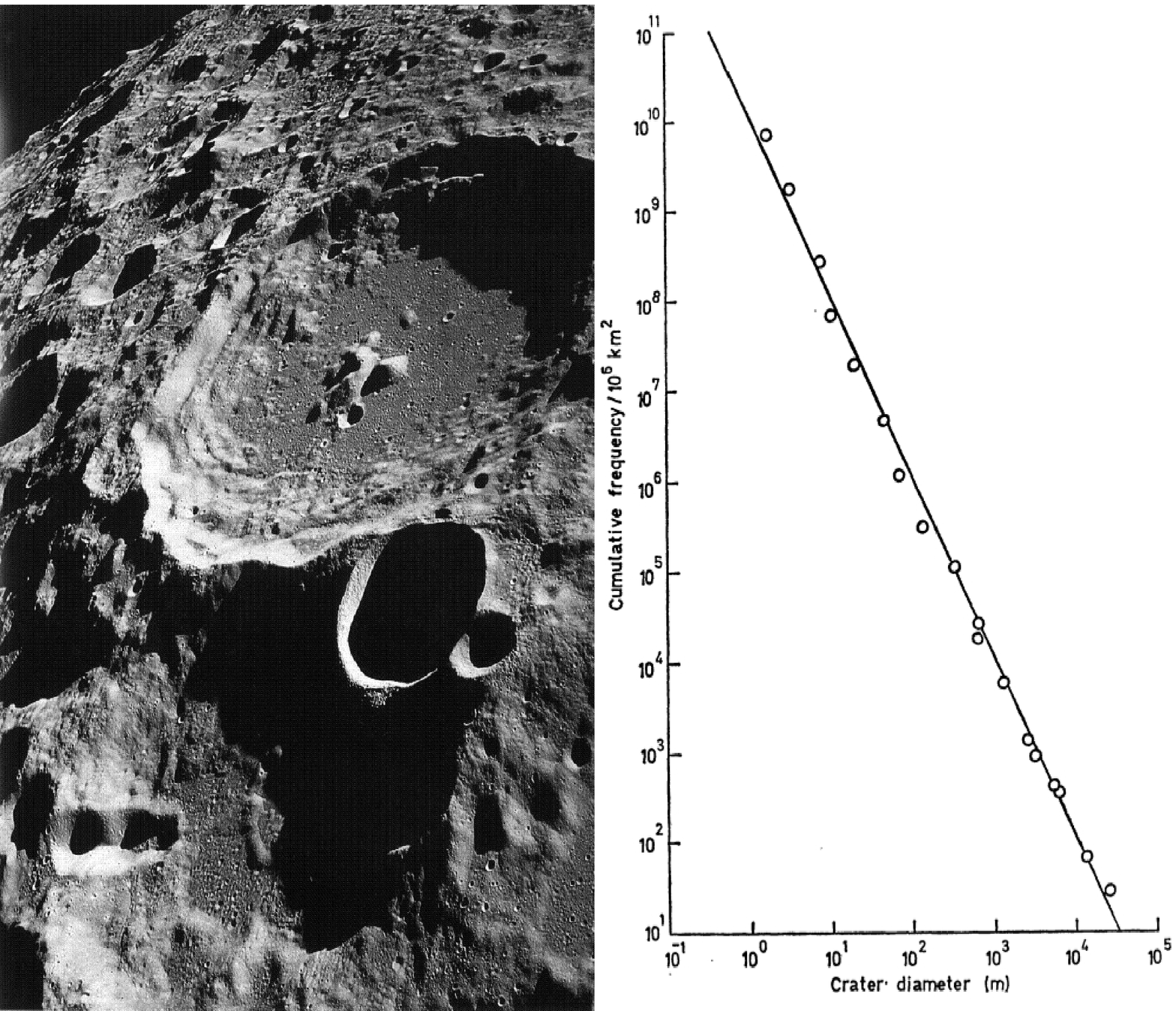}}
\captio{{\sl Left:} The lunar crater Daedalus, about 93 km in diameter,
was photographed by the crew of Apollo 11 as they orbited the Moon in
1969 (NASA photo AS11-44-6611). {\sl Right:} Cumulative frequency
distribution of crater diameters measured from {\sl Ranger 8} in
the lunar {\sl Mare Tranquillitatis} (Cross 1966).}   
\end{figure}

\subsection{Lunar Craters}

The size of lunar craters was measured from pictures recorded with the
lunar orbiters {\sl Ranger 7, 8, 9} by Cross (1966). A size distribution
of 1,600 lunar craters, sampled in the {\sl Mare Tranquillitatis} using
data from {\sl Ranger 8}, within a range of 0.56 to 69,000 m, is shown in
Fig.~13.3, exhibiting a powerlaw distribution ranging over 5 orders of
magnitude with a slope of $\beta \approx 2.0$ for the cumulative 
occurrence frequency distribution, which translates into a powerlaw
slope of $\alpha \approx \beta+1 \approx 3.0$ for the differential occurrence
frequency distribution,
$$
	N(L) \propto L^{-3} \ .
	\eqno(13.31)
$$
This corresponds exactly to our prediction of the scale-free probability
theorem for avalanche events in 3D-space ($S=3$). A similar powerlaw index 
of $\alpha_L=2.75$ was also found for the size distribution of meteorites
and space debris from man-made rockets and satellites (Fig.~3.11 in
Sornette 2004). 

How can we interpret this result? The justification of our scale-free
probability theorem is the fact that the relative probability of 
partitioning a system into smaller parts scales reciprocally
with the volume (Fig.~13.1). The leading theory of lunar crater 
formation is
that their origin was caused by impacts of meteorites, and thus the
size distribution reflects that of the impacting meteors and meteorites, 
which probably were produced by numerous random collisions, similar to the
origin of planets, asteroids, and planetesimals. Both the Moon and the
Earth were subjected to intense bombardment of solar system bodies
between 4.6 and 4.0 billion years ago, which was the final stage of
the sweep-up of debris left over from the formation of the solar system.

An alternative explanation of lunar or terrestrial craters is a
volcanic origin. If we attribute volcanoes with nonlinear energy
dissipation avalanches in a slowly-driven system of stressing planet crust 
motions and build-up of subtectonic lava pressure, volcanic eruptions 
can also be understood as SOC phenomena.

\begin{figure}
 \centerline{\includegraphics[width=0.4\textwidth]{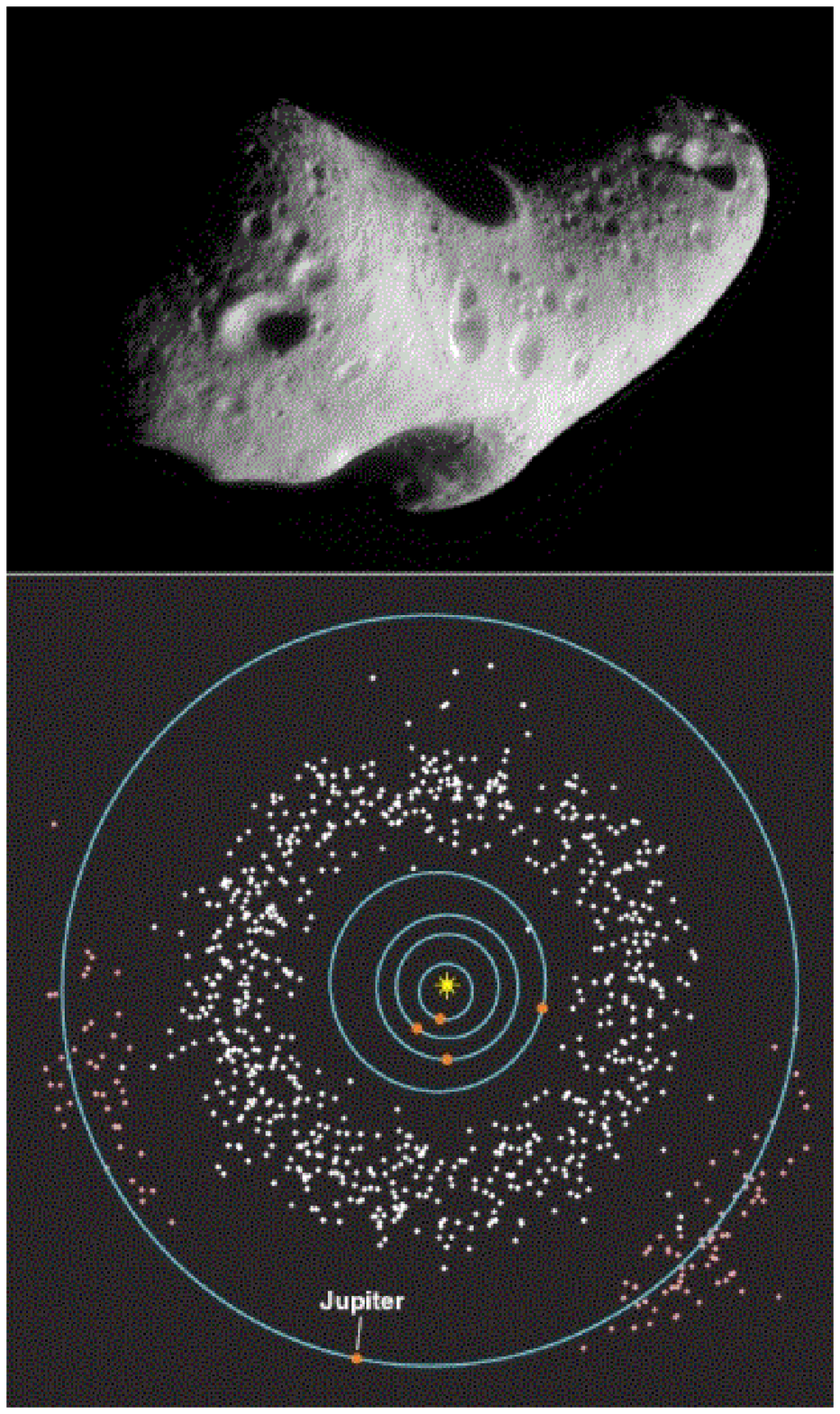}
             \includegraphics[width=0.6\textwidth]{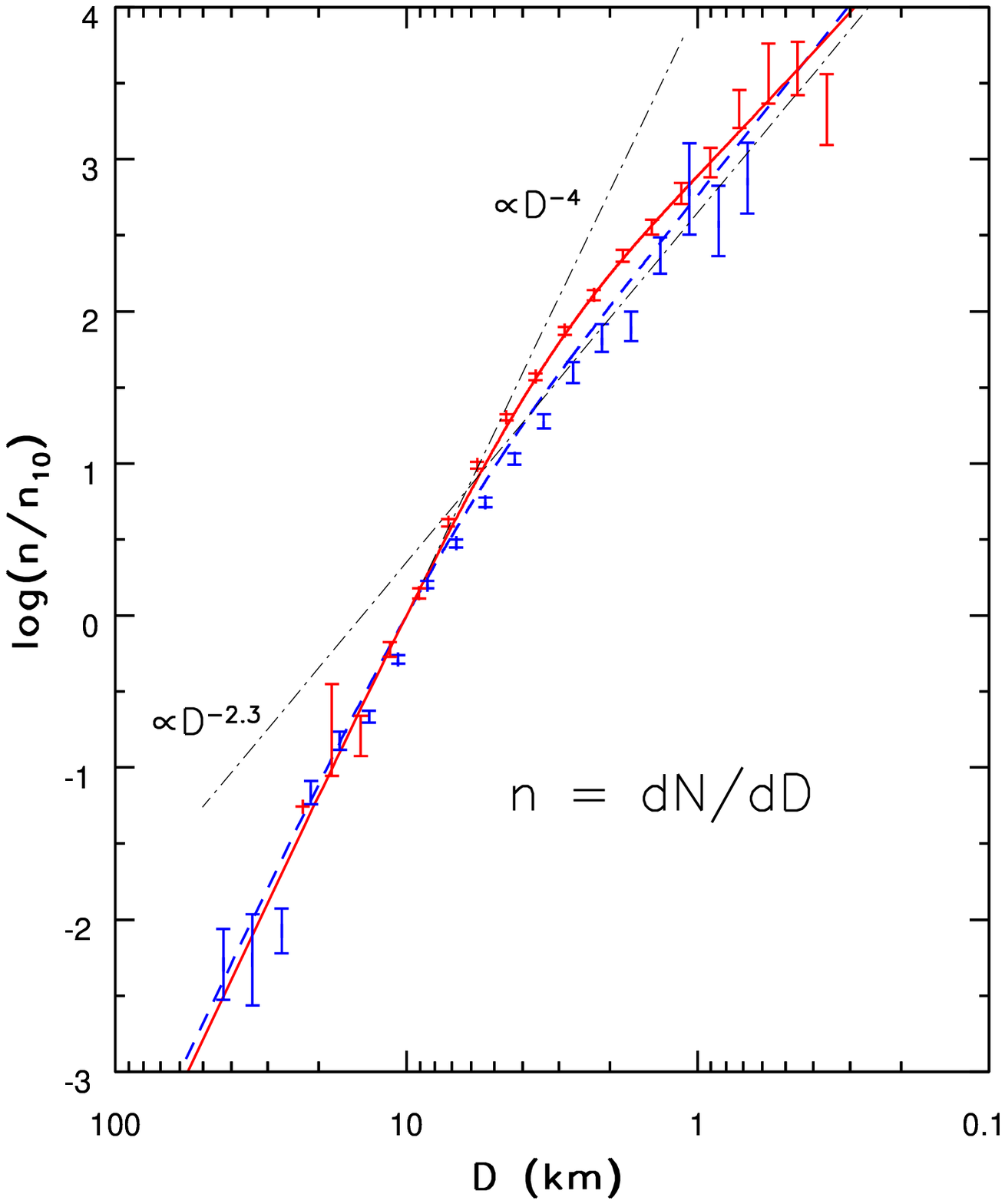}}
\captio{{\sl Left top:} A picture of the near-Earth asteroid {\sl Eros}
with a size of 30 km, pictured by a space probe.
{\sl Left bottom:} The main asteroid belt located between the Jupiter
and Mars orbit. The subgroup of {\sl Trojan asteroids} are leading
and trailing along the Jupiter orbit.
(Courtesy of NASA/Johns Hopkins University Applied Physics Laboratory).
{\sl Right:} Differential size distribution of asteroids observed in the
{\sl Sloan Digital Sky Survey} collaboration
(Ivezic et al.~2001).}  
\end{figure}

\subsection{Asteroid Belt}

The asteroid size distribution has been studied in the {\sl Palomar Leiden
Survey} (Van Houten et al.~1970) and {\sl Spacewatch Surveys}
(Jedicke and Metcalfe 1998), where a power law of 
$N^{cum}(>L) \propto L^{-1.8}$
was found for the cumulative size distribution of larger asteroids
($L > 5$ km), which corresponds to a differential powerlaw slope of
$\alpha_L \approx 2.8$. In a {\sl Sloan Digital Sky Survey} collaboration
(Fig.~13.4, right), a broken powerlaw was found with 
$N(L) \propto L^{-2.3}$ for large asteroids
(5-50 km) and $N(L) \propto L^{-4}$ for smaller asteroids (0.5-5 km)
(Ivezic et al.~2001). In the {\sl Subaru Main-Belt Asteroid Survey},
a cumulative size distribution $N^{cum}(>L) \propto L^{-1.29\pm0.02}$
was found for small asteroids with $L\approx 0.6-1.0$ km
(Yoshida et al.~2003; Yoshida and Nakamura 2007),    
which corresponds to a differential powerlaw slope of 
$\alpha_L \approx 2.3$. Thus, the observed range $\alpha_L \approx 2.3-4.0$
of the powerlaw slopes of length scales $L$ is centered around the 
theoretically expected value of $\alpha_L=3.0$, predicted by our 
scale-free probability theorem.  

The origin of asteroids is thought to be a left-over distribution of
planetesimals during the formation of planets, which were either too small 
to form bigger planets by self-gravitation, or orbited in an unstable
region of the solar system where Mars and Jupiter constantly cause
gravitational disturbances that prevented the formation of another
planet. Thus, the final distribution of the asteroid belt is likely
to be influenced by both the primordial distribution of the solar system
as well as by recent collisions and further fragmentation of planetesimals.
The collisional fragmentation process can be considered as a mechanical
instability that occurs in a multi-body gravitational field. 
The collisional process is self-organizing in the sense that the
N-body celestial mechanics keeps the structure of the asteroid belt 
more or less stable, despite of the combined effects of self-gravity, 
gravitational disturbances, collisions, depletions, and captures of 
incoming new bodies. The quasi-stability of the asteroid belt warrants 
the critical threshold in form of a finite collision probability 
maintained by the proximity of the co-orbiting asteroid bodies.

\begin{figure}
\centerline{\includegraphics[width=1.\textwidth]{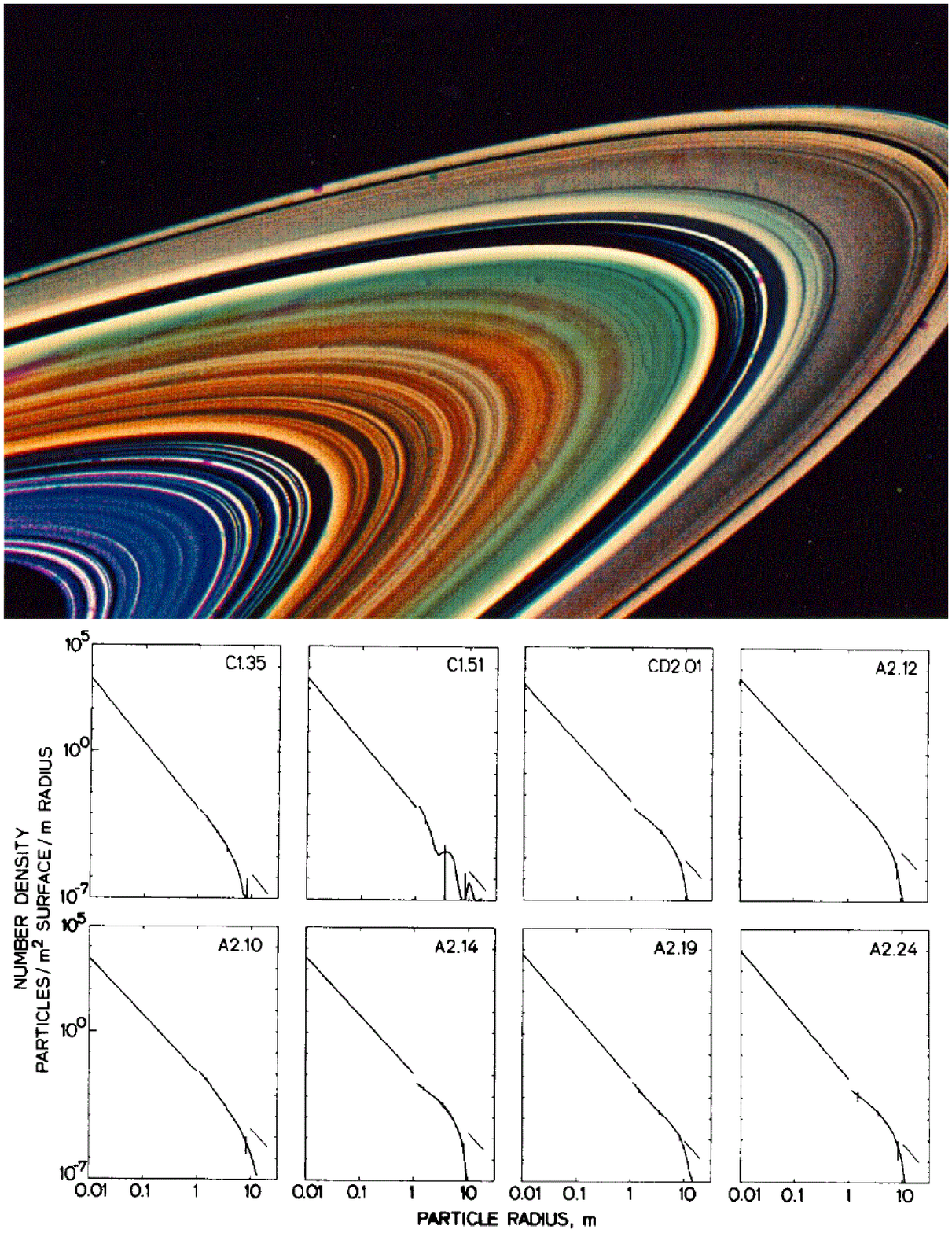}}
\captio{{\sl Top:} Saturn's rings A, B, C, and the Cassini
division, photographed by the {\sl Cassini} spacecraft
(credit: NASA, JPL, Space Science Institute). 
{\sl Bottom:} Measurements of the particle size distribution
functions for 8 ring regions with {\sl Voyager I} radio occultation
measurements (Ring C: C1.35, C1.51; Cassini division:
CD2.01; Ring A: A2.12, A2.10, A2.14, A2.19, A2.24). The slopes of the
fitted powerlaw functions in these 8 regions are: $\alpha_L=$
3.11, 3.05, 2.79, 2.74, 2.70, 2.75, 2.93, 3.03. The range of
particle sizes is $L=0.01-10$ m (Zebker et al.~1985).}
\end{figure}

\subsection{Saturn Ring}

The distribution of particle sizes in Saturn's ring was determined
with radio occultation observations using data from the {\sl Voyager 1}
spacecraft and a scattering model, which exhibited a powerlaw distribution 
of $N(r) \propto r^{-3}$ (Fig.~13.5) in the range of
1 mm $<$ r $<$ 20 m (Zebker et al.~1985; French and Nicholson 2000).
The size distribution revealed slightly different powerlaw 
slopes in each ring zone, e.g., $\alpha_L=2.74-3.03$ for ring A,
$\alpha_L=2.79$ for the Cassini division, or 
$\alpha_L=3.05-3.22$ for ring C (Zebker et al.~1985).
These results, again, are consistent with a fragmentation process that
obeys the scale-free probability theorem, similar to the distribution 
of sizes of asteroids and lunar craters, and predicts a size
distribution of $N(L) \propto L^{-3}$. The conclusion that Saturn's 
ring particles are formed from the (collisional) breakup of larger
particles, rather than from original condensation as small particles,
was already raised earlier (Greenberg et al.~1977). 

The Saturn ring, which consists of particles ranging in size from
1 cm to 10 m, is located at a distance of 7,000-80,000 km above 
Saturn's equator, and has a total mass of $3 \times 10^{19}$ kg, 
just about a little less than the moon Mimas. The origin of the ring 
is believed to come either from leftover material of the formation of
Saturn itself, or from the tidal disruptions of a former moon.
The celestial mechanics of the Saturn rings is quite complex, 
revealing numerous gaps in orbits that have a harmonic ratio
in their period with one of the 62 (confirmed) moons (with 13
moons having a size larger than 50 km). Similar to the asteroid belt,
the Saturn ring can be considered as a self-organizing system in the
sense that the gravity of Saturn and the gravitational disturbances
caused by Saturn's moons keep the ring quasi-stable, which provides
a critical threshold rate for collisional encounters due to the proximity
of the co-orbiting ring particles.

\begin{figure}
\centerline{\includegraphics[width=1.\textwidth]{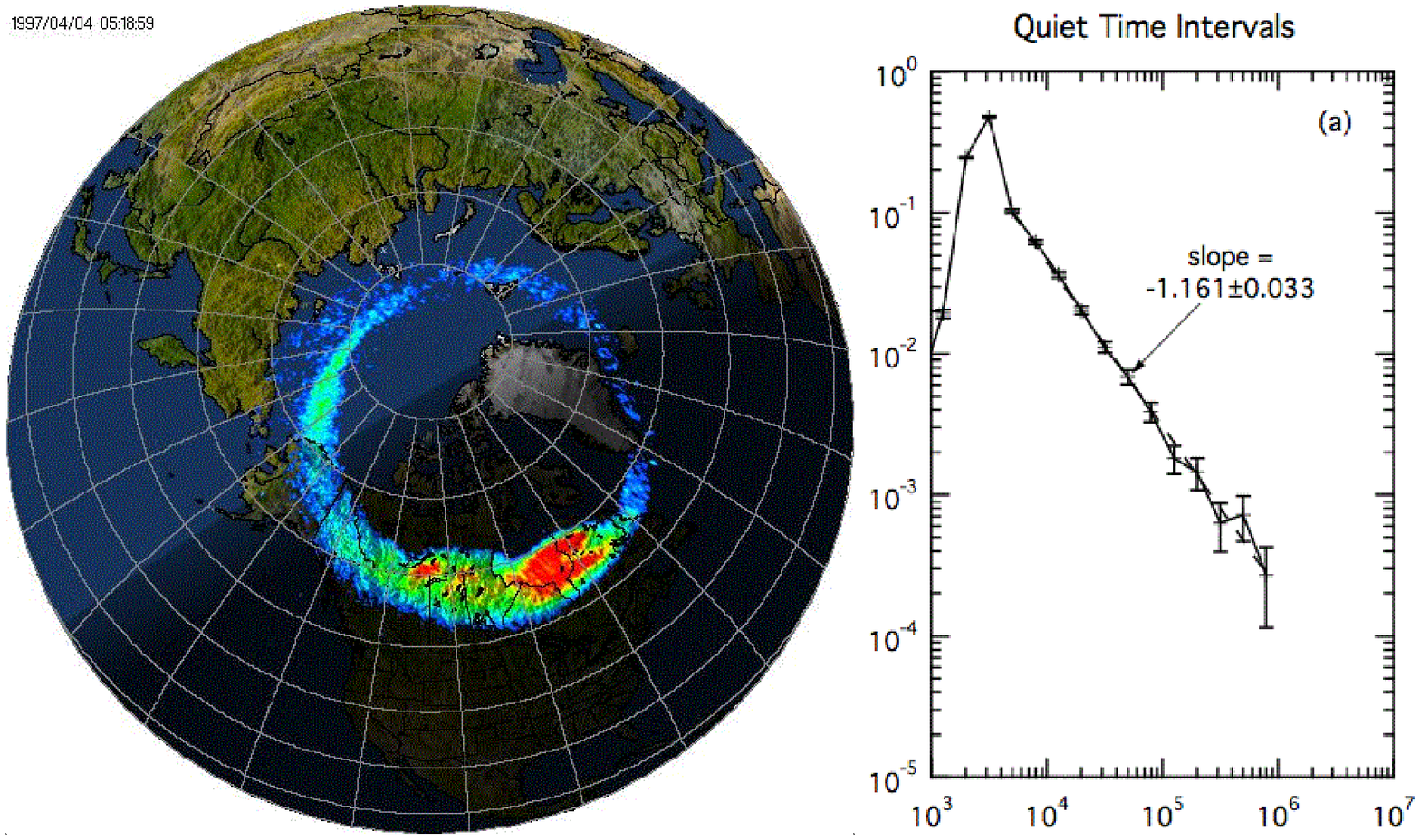}}
\captio{{\sl Left:} Global image of the auroral oval observed by the 
Ultraviolet Imager (UVI) onboard the NASA satellite ``Polar'' on 
April 4, 1997 at 0519 UT, projected onto an Earth map (credit: NASA, 
Polar/UVI Team, George Parks).
{\sl Right:} Occurrence rate frequency distributions of auroral blobs
as a function of the area (in units of square kilometers) during 
substorm-quiet time intervals, recorded with {\sl Polar UVI} during
Jan 1-31, 1997 (Lui et al.~2000).}   
\end{figure}

\subsection{Magnetospheric Substorms and Auroras}

The size distribution of auroral areas have been measured with the UV Imager 
of the {\sl Polar} spacecraft, which exhibits a powerlaw-like distribution
with a slope of $\alpha_A=1.21\pm0.08$ during active substorm time
intervals, and $\alpha_A=1.16\pm0.03$ during quiescent time intervals
(Fig.~13.6; Lui et al.~2000). The corresponding energy flux or power output 
$P$ of auroral regions was derived to have a powerlaw slope of
$\alpha_P=1.05\pm0.08$ during active substorm time intervals, and
$\alpha_P=1.00\pm0.02$ during quiescent time intervals (Lui et al.~2000).
These powerlaw slopes are significantly flatter than predicted by our 
FD-SOC model, i.e., $\alpha_A=2.33$ and $\alpha_P=1.67$ for 3D phenomena 
($S=3, D_2 \approx 1.5$). Although Lui et al.~(2000) interpret auroras
as a SOC phenomenon, the observed powerlaw slopes are far
out of the range observed and predicted for other SOC phenomena.
Thus, either the sampled distributions are incomplete, they underestimate 
the areas systematically for smaller events, or SOC models are not 
applicable for these events.

Plasma flows in the magnetotail plasma with speeds $v \ge 400$ km s$^{-1}$
were found to have a powerlaw distribution of durations $T$, with 
$N(T) \propto T^{1.59\pm0.07}$ (Angelopoulos et al.~1999), which is not 
too far off our theoretical prediction (with $\alpha_T =2.0$), given
the relatively small powerlaw range of only $\approx 1.5$ decades. 
Also the size distributions of the durations of AE index ($\alpha=1.24$;
Takalo 1993; Takalo et al.~1999) and AU index ($\alpha=1.3$;
Freeman et al.~2000; Chapman and Watkins 2001) were found to be much
flatter than predicted by our SOC model.

Electron bursts in the outer radiation belt (at $4-8$ L-shell distances),
which may be modulated by fluctuations of the solar wind, were found to
have powerlaw distributions with slopes of $\alpha_P=1.5-2.1$ and were
interpreted as SOC phenomena (Crosby et al.~2005), which have slopes that
are quite consistent with our SOC model ($\alpha_P=1.67$). The solar wind 
is thought to be the source of these energetic electrons, although the 
solar wind velocity frequency distributions were found to exhibit 
significant deviations from simple powerlaws (Crosby et al.~2005). 

\begin{figure}
\centerline{\includegraphics[width=1.0\textwidth]{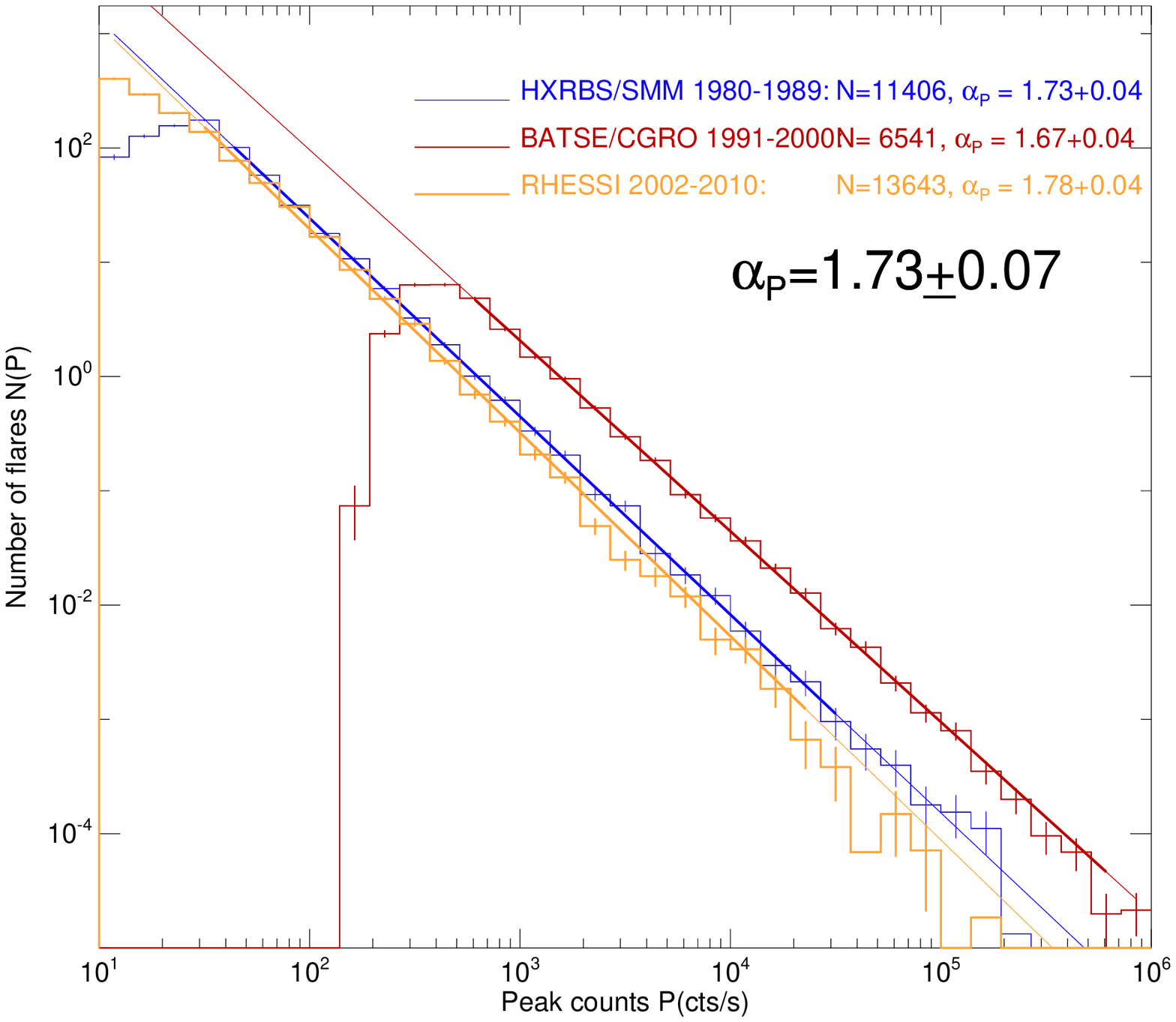}}
\captio{Occurrence frequency distributions of hard X-ray peak count rates
$P$ [cts s$^{-1}$] observed with HXRBS/SMM (1980 -- 1989), BATSE (1991 -- 2000),
and RHESSI (2002 -- 2010), with powerlaw fits. An average pre-flare
background of $40$ [cts s$^{-1}$] was subtracted from the HXRBS
count rates. Note that BATSE/CGRO has larger detector areas, and thus
records higher count rates (Aschwanden 2011b).}
\end{figure}

\begin{figure}
\centerline{\includegraphics[width=1.0\textwidth]{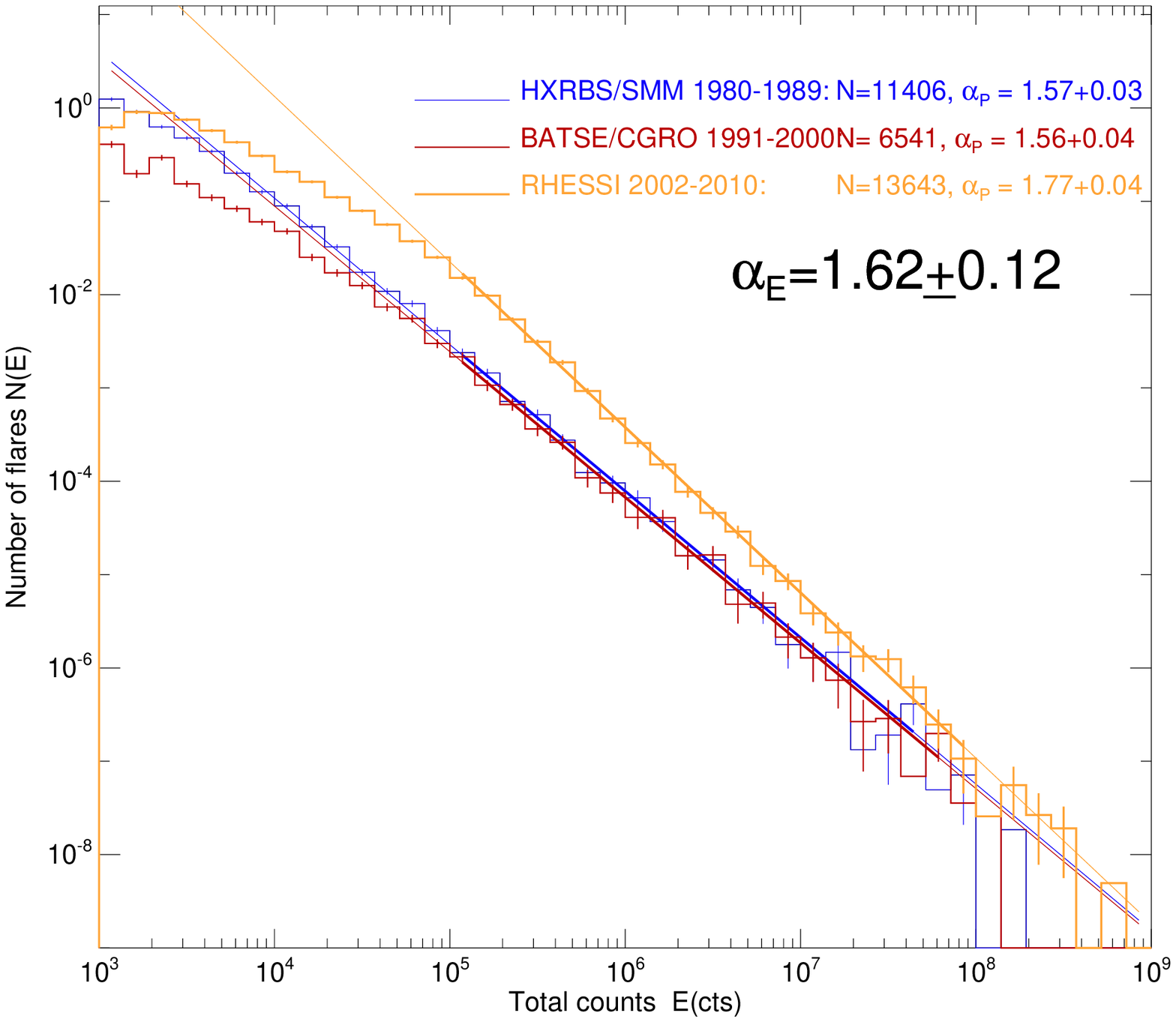}}
\captio{Occurrence frequency distributions of hard X-ray total counts
or fluence $E$ [cts] observed with HXRBS/SMM (1980 -- 1989),
BATSE (1991 -- 2000), and RHESSI (2002 -- 2010), with powerlaw fits.
An average pre-flare background of $40$ cts s$^{-1}$ multiplied with
the flare duration was subtracted in the total counts of HXRBS
(Aschwanden 2011b).}
\end{figure}

\begin{figure}
\centerline{\includegraphics[width=1.0\textwidth]{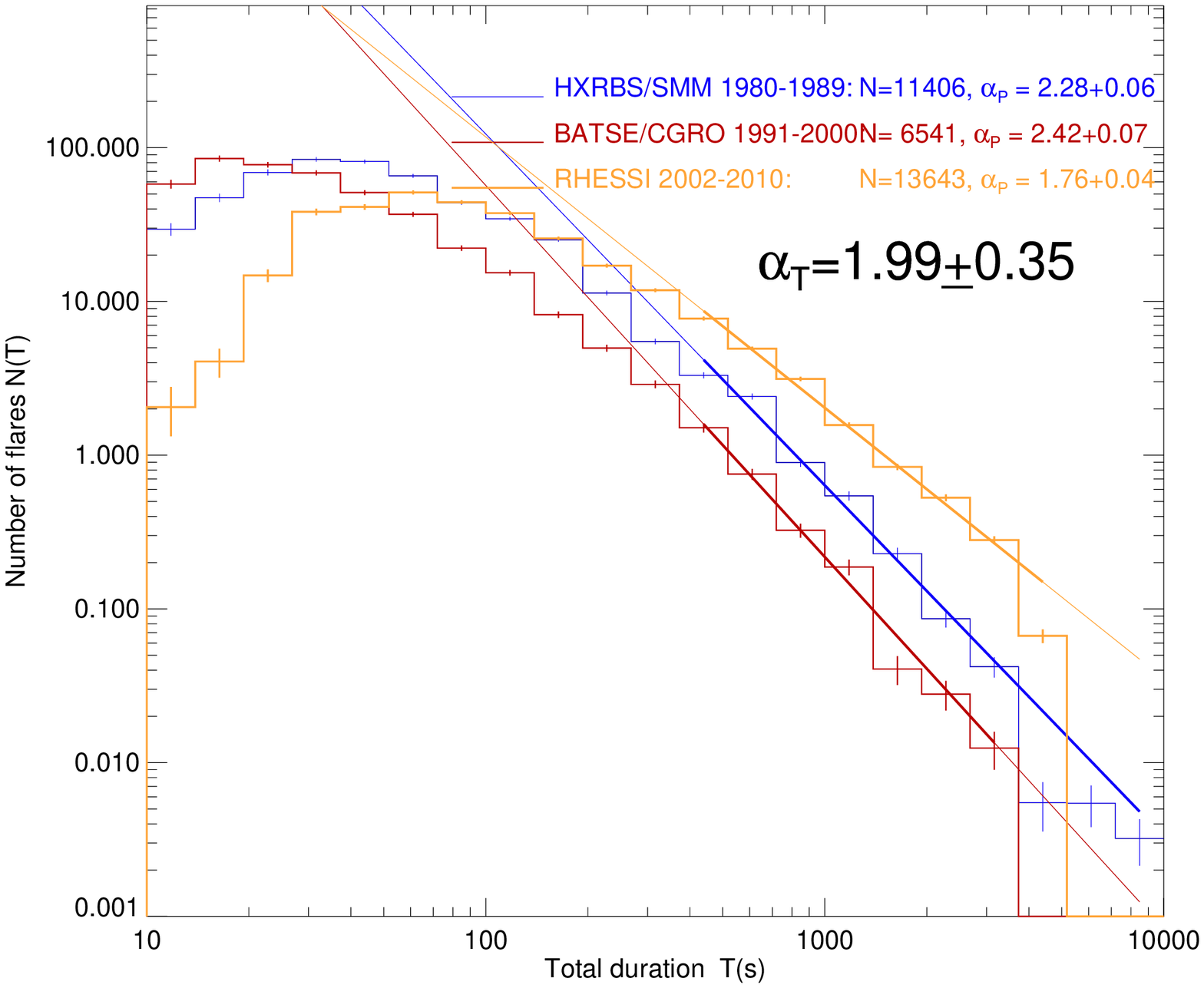}}
\captio{Occurrence frequency distributions of hard X-ray flare durations
$T$ [s] observed with HXRBS/SMM (1980 -- 1989), BATSE (1991 -- 2000),
and RHESSI (2002 -- 2010) with powerlaw fits. The flare durations for RHESSI
were estimated from the time difference between the start and peak time,
because RHESSI flare durations were determined at a lower energy of 12 keV
(compared with 25 keV for HXRBS and BATSE), where thermal emission prolonges
the nonthermal flare duration (Aschwanden 2011b).}
\end{figure}

\subsection{Solar Flares}

Solar flares are the best studied SOC phenomena in astrophysics.
The impulsive energy release associated with solar flares, which can be 
observed in virtually all wavelengths, from gamma-rays, hard
X-rays, soft X-rays, EUV, white-light, infrared, to radio wavelengths,
has been interpreted as a SOC phenomen from early on 
(Lu and Hamilton 1991). Large datasets with $n \approx 10^4-10^5$
events sampled over up to eight orders of magnitude in energy provide
the necessary statistics to determine accurate slopes of the 
observed powerlaw-like size distributions. However, major challenges
exist still in the elimination of sampling biases in incomplete
event sets, the understanding and modeling of powerlaw slopes in
different wavelengths in terms of the underlying physical scaling laws,
and the automated determination of geometric parameters for large
event datasets. A detailed account of observational results sorted
into different wavelength regimes is given in Aschwanden (2012a;
chapters 7 and 8). We summarize the results of powerlaw slopes
observed in size distributions of various SOC parameters in Table 13.2,
selecting mostly representative examples with large datasets from 
different instruments and wavelength regimes.

Size distributions of peak fluxes $P$ (Fig.~13.7), time-integrated 
fluxes or fluences $E$ (Fig.~13.8), and flare durations $T$ (Fig.~13.9), 
have been measured for energies 
$>25$ keV in hard X-ray wavelengths with instruments on the spacecraft
ISEE-3, SMM, CGRO, and RHESSI, in soft X-ray wavelengths 
with Yohkoh and GOES, and in EUV with SOHO/EIT, TRACE, and AIA/SDO.
Most of the observed powerlaw slopes were measured close to the
theoretical predictions, i.e., $\alpha_T=2.0$, $\alpha_P=1.67$, and
$\alpha_E=1.5$ (Table 13.2), which is consistent with a dimensionality of 
$S=3$, a mean fractal dimension of $D_S \approx (1+S)/2 = 2.0$, an
energy-volume scaling index of $\gamma \approx 1$, and a diffusion
power index of $\beta \approx 1$ (Eq.~13.17 and 13.19). 

The measurements in soft X-rays and hard X-rays are all made in 
broadband energy and wavelength ranges, and thus are least biased 
regarding a complete sampling of all energy and temperature ranges.
The probably most controversial measurements have been made for the
smallest flares, also called nanoflares, which have typical 
temperatures of $T\approx 1-2$ MK and originate in small loops
that barely stick out of the transition region. Since EUV measurements
with SOHO/EIT, TRACE, and AIA/SDO are all made with narrowband 
temperature filters, the inferred thermal energies essentially
scale with the flare area (Eq.~13.27) or volume (Eq~13.26), for which
we predict powerlaw slopes of $\alpha_{th,A}\approx \alpha_A \approx 2.3$ and
$\alpha_{th,V} \approx \alpha_V \approx 2.0$, which are significantly steeper
than what is predicted for thermal energies sampled with broadband
instruments, i.e., $\alpha_{th} \approx 1.5$. This sampling bias has
resulted into a controversy whether nanoflares dominate coronal heating,
because a powerlaw slope steeper than the critical value of 2 indicates
that the energy integral diverges for the smallest events, as pointed
out early on (Hudson 1991). Synthesizing measurements from narrowband
EUV instruments with broadband soft X-ray instruments, as well as taking
the fractal geometry of flare structures into account, however, could
reconcile the size distribution for nanoflares with that of large flares
with a corrected value that is close to the theoretical prediction
of $\alpha_E \approx 1.5$ (Fig.~13.10; Aschwanden and Parnell 2002). 

Problematic are also the measurements of flare durations $T$ for several
reasons, such as the limited range of durations over which a powerlaw
can be fitted, the ambiguity of separating overlapping long-duration
flares, and the solar cycle dependence. While the event overlap problem
is not severe during the solar minimum, where a slope close to the
theoretically predicted value of $\alpha_T=2.0$ is measured, the
flare event pile-up bias becomes very severe during the solar maximum,
producing powerlaw slopes of up to $\alpha_T \lapprox 5$ (Aschwanden
and Freeland 2012). Also the powerlaw slope of hard X-ray peak counts
$\alpha_P$ appears to reveal a solar cycle dependence due to a similar
effect (Bai 1993, Biesecker 1994, Aschwanden 2011b). 

\begin{figure}
\centerline{\includegraphics[width=1.0\textwidth]{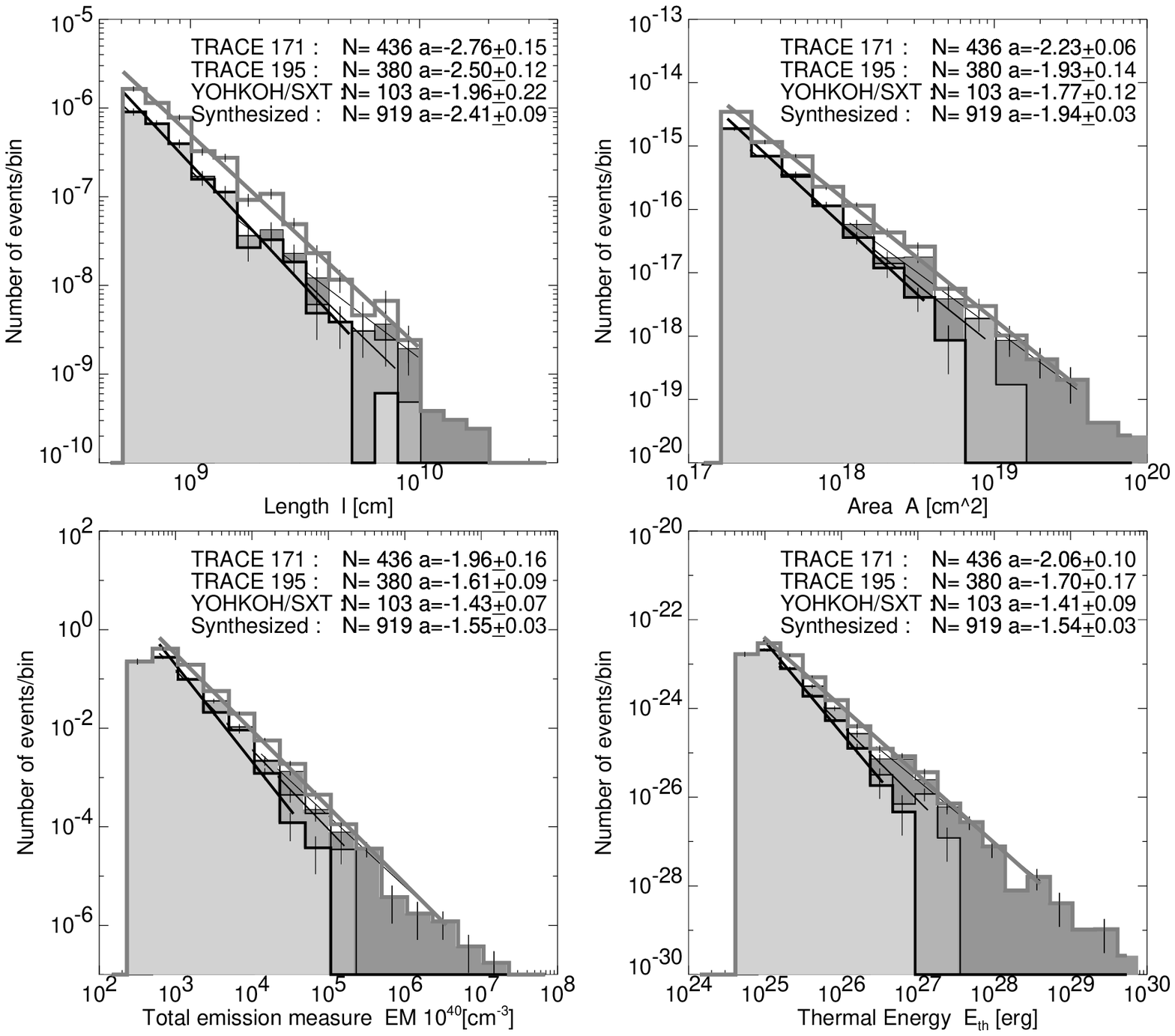}}
\captio{Synthesized frequency distributions from all three wavebands
(TRACE 171 \ang, 195 \ang , and Yohkoh/SXT AlMg) (grey histograms), along with
the separate distributions from each waveband (in greyscales). Each of the
distributions is fitted with a powerlaw, with the slope values and formal
fit errors given in each panel. The four panels represent the four parameters
of length $L$, area $A$, total emission measure $EM$ (which is proportional
to the peak flux $P$), and the thermal energy $E_{th}$ 
(Aschwanden and Parnell 2002).}  
\end{figure}

\begin{figure}
\centerline{\includegraphics[width=1.0\textwidth]{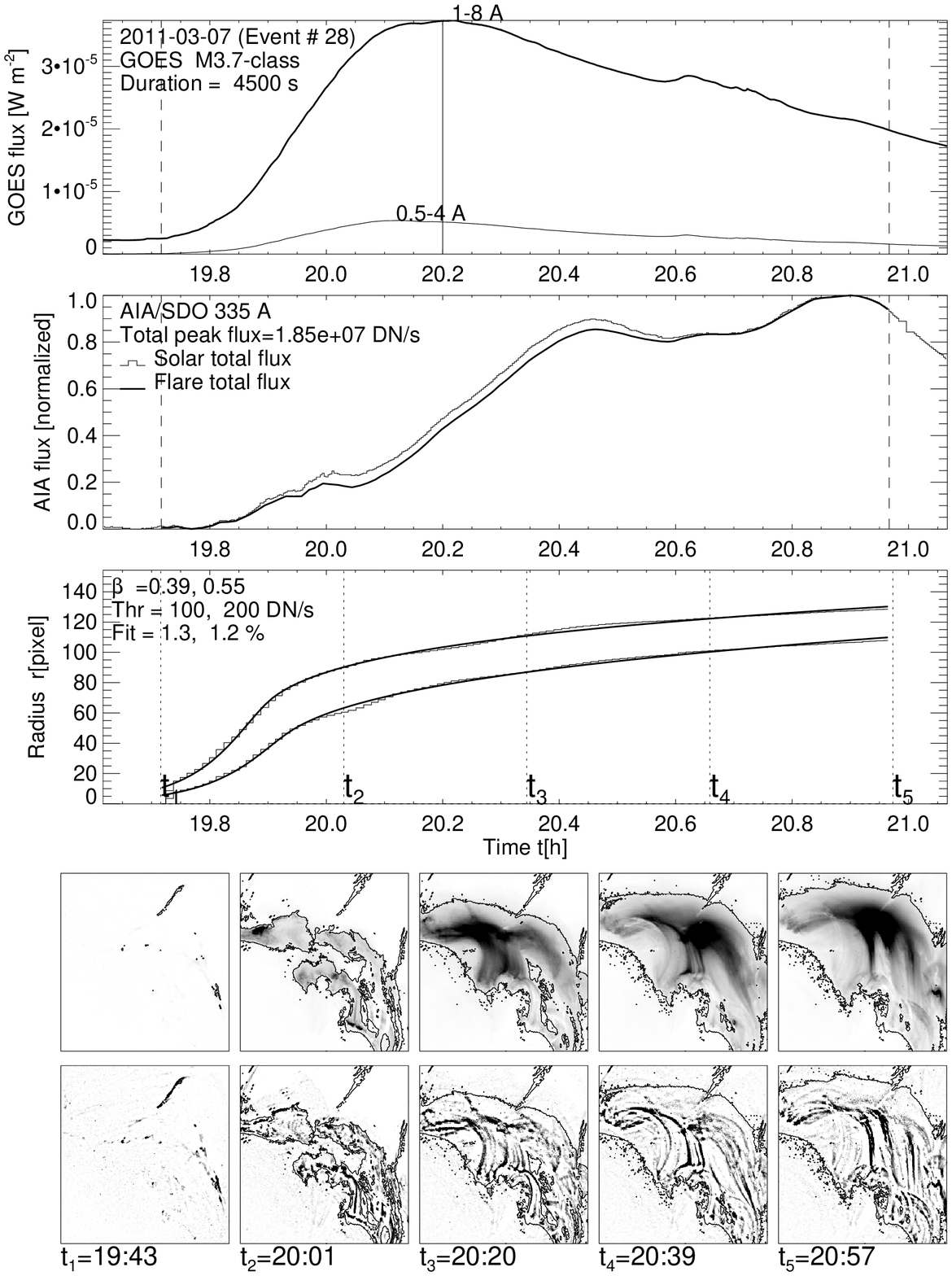}}
\captio{A solar flare event is observed on 2011 Mar 7, 19:43-20:58 UT,
with AIA/SDO 335 \ang , with GOES time profiles (top panel), the
total EUV 335 \ang\ flux (second panel), the spatio-temporal evolution 
of the radius
$r(t)=\sqrt{A(t)/\pi}$ of the time-integrated flare area $A(t)$
for two thresholds, $F_{thresh}=100, 200$ DN/s (third panel; histogrammed),
fitted with the anomalous diffusion model (third panel; solid curve),
and 5 snapshots of the baseline-subtracted flux (fourth row) and
highpass-filtered flux (bottom row), with the threshold flux
$F_{thresh}=100$ DN/s shown as contour (Aschwanden 2012b).}
\end{figure}

\begin{figure}
\centerline{\includegraphics[width=1.0\textwidth]{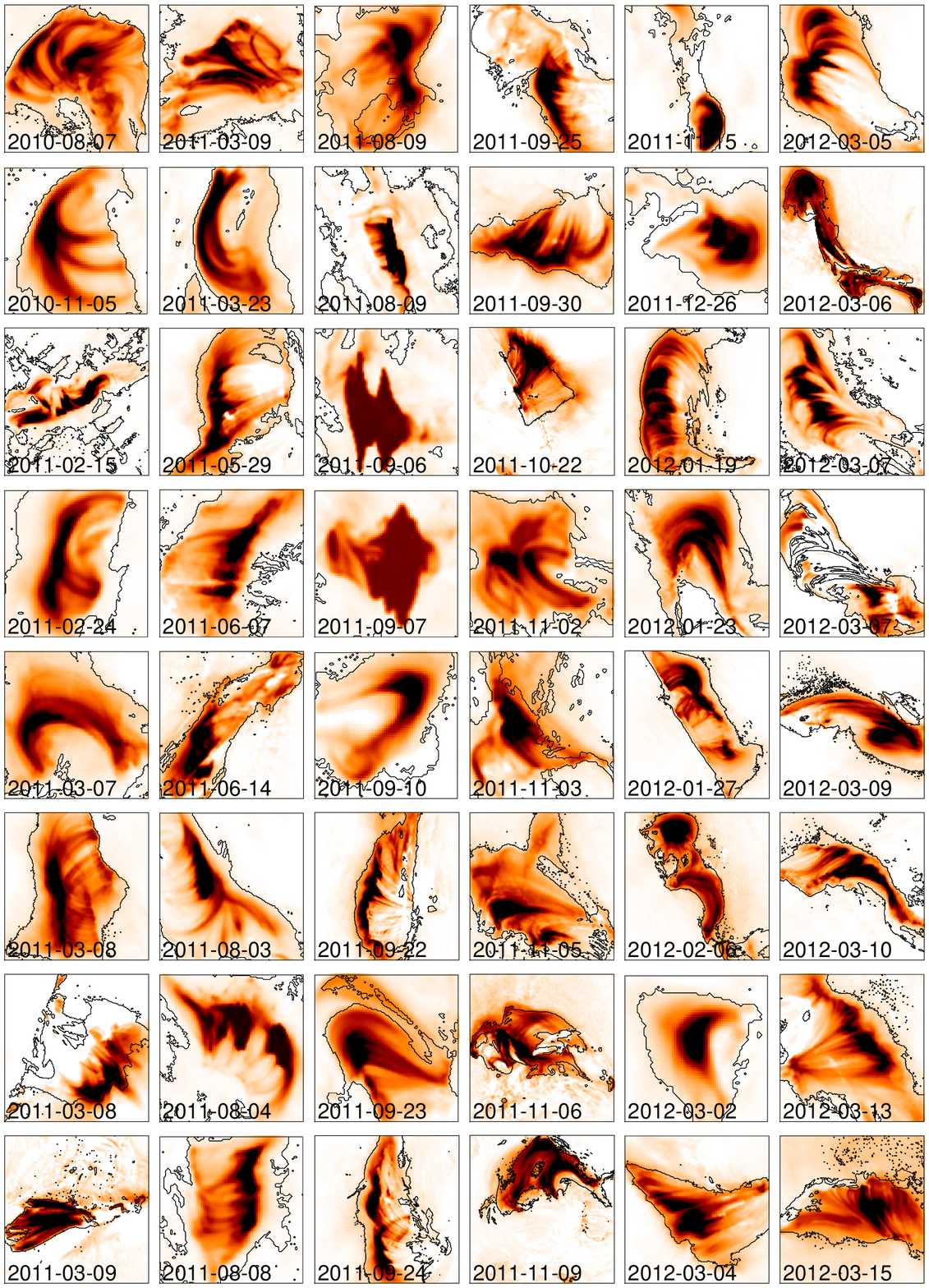}}
\captio{A selection of 48 solar GOES M and X-class flares observed 
with AIA/SDO at a wavelength of 335 \ang\ during 2010-2012. The
contour shows the time-integrated flare area and the color-scale
represents the intensity at 335 \ang . Note the complex spatial
patterns that ressemble to the fractal avalanche patterns of
cellular automaton simulations.}
\end{figure}

The least explored size distributions of solar flares are the length
scale $L$ and area $A$ size distributions. Relatively small samples
of flare areas have been measured for EUV nanoflares 
(Aschwanden et al.~2000, Aschwanden and Parnell 2002) 
and for the largest M and X-class flares (Aschwanden 2012b).
Since the measurement of these parameters provides a direct test of
the scale-free probability theorem (Eq.~13.1), without depending on
any other physical parameter or model assumption, priority should be
given to such measurements. Existing measurements have large error bars
in the powerlaw slope due to the small number of analyzed events,
but are largely consistent with the theoretical prediction of
$\alpha_L=3.0$ and $\alpha_A=2.33$, calculated for 3D 
SOC avalanches ($S=3$) with a 2D (area $S=2$) fractal dimension of 
$D_2 \approx (1+S)/2 = 1.5$. 

Besides the scale-free probability theorem, a second pillar of our
FD-SOC model is the fractal-diffusive spatio-temporal relationship
$L(t) \propto \kappa T^{\beta/2}$ (Eq.~13.8), which has been
recently tested for a set of the 155 largest (GOES M and X-class)
flare events (Aschwanden 2012b). An example of such a measurement
of the spatio-temporal evolution is shown in Fig.~13.11, which
exemplifies the spatial complexity of a large cluster of subsequent
magnetic reconnection events that form a fractal volume filled with
postflare loops. Examples of another 48 large flares (GOES M and 
X-class) are shown in Fig.~13.12.
Interestingly, the statistics of these 155 largest
flares revealed a sub-diffusive regime ($\beta = 0.53 \pm 0.27$),
while classical diffusion appears to be an upper limit. The 
diffusive characteristics measured in solar flares is consistent
with the FD-SOC model, as it is was also found to be consistent with
cellular automaton SOC simulations (Aschwanden 2012a). 

Radio bursts are produced in solar flares most frequently either by 
gyrosynchrotron emission of relativistic electrons that have been
accelerated in magnetic reconnection regions, or by electron beams
that escape along magnetic field lines in upward-direction. Both
types of radio bursts (microwave bursts and type III bursts) occur as a
consequence of a plasma instability, and thus represent a highly 
nonlinear energy dissipation process that is typical for SOC processes.

Somewhat ouf the predicted range are solar energetic particles, which
have rather flat size distributions $N(P)$ of the peak counts, 
typically in the range of $\alpha_P \approx 1.1-1.4$. However, since
these are all high-energy particle events ($>10$ MeV protons and
$>3$ MeV electrons), we suspect that only the largest flares produce
such high energies, and thus the sample is biased towards the largest
flare events, which explains the flatter powerlaw slopes as a 
consequence of missing weaker events. 

\begin{figure}
\centerline{\includegraphics[width=1.0\textwidth]{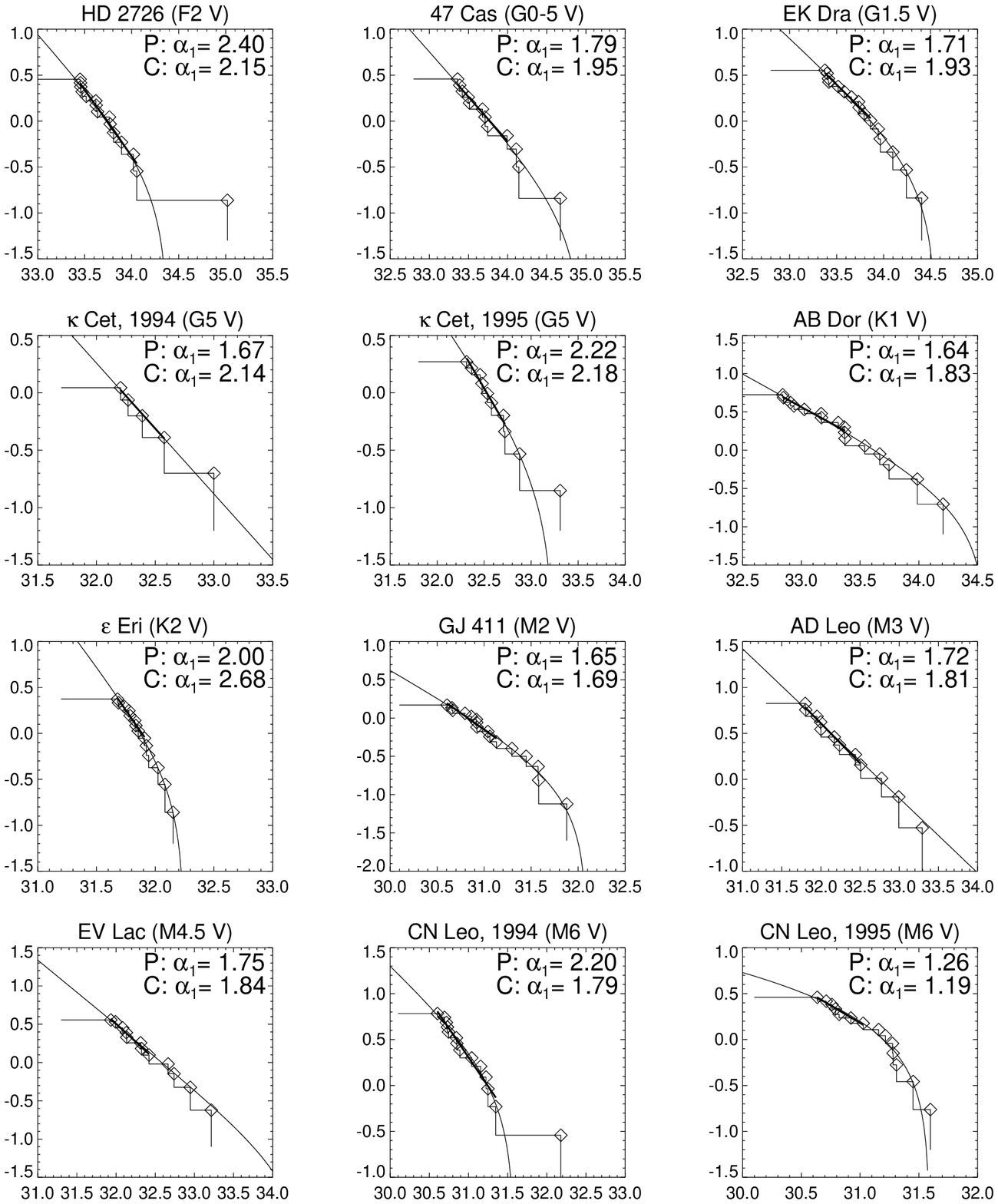}}
\captio{Cumulative frequency distributions of flare energies (total counts)
observed for 12 cool (type F to M) stars with EUVE (Audard et al.~2000). 
The flare events are marked with diamonds, fitted with a powerlaw fit in
the lower half (P; thick line), and fitted with a cumulative frequency
distribution (C; curved function).}
\end{figure}

\subsection{Stellar Flares}

Impulsive flaring with rapid increases in the brightness in UV or EUV has
been observed  for a number of so-called flare stars, such as AD Leo, 
AB Dor, YZ Cmi, EK Dra, or $\epsilon$ Eri. These types of stars include
cool M dwarfs, brown dwarfs, A-type stars, giants, and binaries in the
Hertzsprung-Russell diagram. Most of these stars are believed to have
hot soft X-ray emitting coronae, similar to our Sun (a G5 star), and
thus magnetic reconnection processes are believed to operate in a similar
way as on our Sun. 

However, what is different, is that the soft X-ray
emission is several orders of magnitude stronger than from our Sun, if we
put the stars into the same distance, and thus we expect an observational
selection bias towards the largest possible flares. Another difference in
flare statistics is that the meagre observational time allocation in the
order of a few hours per star reveals only very few detectable flare
events, in the order of 5-15 events per observed star. A consequence
of this small-number statistics is that we cannot determine a powerlaw
slope from a log(N)-log(S) histogram, as we do for larger statistics
(of at least $10^2$ up to $10^5$ events in solar flare data sets), 
but need to resort to 
rank-order plots, which correspond to the inverse distribution of
cumulative occurrence frequency distributions. So, in principle, 
an inverse rank-order diagram can be plotted as shown in Fig.~13.13,
which shows the logarithmic rank versus the flare energy (or total 
counts) for each star, from which the powerlaw slope can be determined.
However, if we deal with cumulative size distributions, we have also to
be aware of the drop-off that results at the upper end of the distribution
due to the missing part in the powerlaw differential occurrence frequency
distribution above the largest event. Thus, while a straight powerlaw
function with slope $\alpha$ can be fitted to a differential frequency 
distribution $N(E)$, 
$$	
	N(E) \propto E^{-\alpha} \ ,
        \eqno(13.32)
$$
the following function need to be fitted to the
cumulative distribution $N_{cum}(>E)$ (Aschwanden 2011a, section 7), 
$$
        N_{cum}(>E) = n {\int_E^{E_{max}} N(E') dE'
                        \over \int_{E_{min}}^{E_{max}} N(E') dE'}
                    = n {\int_E^{E_{max}} E'^{-\alpha} dE'
                         \over \int_{E_{min}}^{E_{max}} E'^{-\alpha} dE'}
                    = n {( E^{1-\alpha}       - E_{max}^{1-\alpha} )\over
                         ( E_{min}^{1-\alpha} - E_{max}^{1-\alpha} )} \ .
        \eqno(13.33)
$$
The fit of this function to the cumulative distribution is shown in
Fig.~13.13 for a set of 12 flare stars, and the resulting values for the
powerlaw slope $\alpha_E$ of the inferred differential occurrence frequency 
distributions (labeled with the letter $C$ in Fig.~13.13). For comparison, 
we fit also a straight powerlaw with slope $\beta = \alpha-1$ to the lower 
half of the cumulative distribution, which is less effected by the upper 
cutoff (labeled with the letter $P$ in Fig.~13.13). In Table 13.2 we
summarize the means and standard deviations of the powerlaw slopes of
flare energies observed on flare stars (see individual values in Table 7.7
of Aschwanden 2011a), which have been reported based on various other
methods used by the authors, $\alpha_E=2.17\pm0.25$. Fitting only the
lower half of the distribution functions we find a significantly lower
value of $\alpha_E=1.85\pm0.31$, or by fitting Eq.~(13.33) we find a
similar range of $\alpha_E=1.93\pm0.35$. This subtle difference in the
determination of the powerlaw slope is essential, because it discriminates
whether the total energy radiated during stellar flares is dominated by the
largest flares (if $\alpha_E < 2.0$) or by nanoflares (if $\alpha_E > 2.0$; 
Hudson 1991). At this point it is not clear whether the difference 
in the powerlaw slopes obtained for stellar versus solar flares is due to 
a methodical problem of small-number statistics, or due to a sampling bias
for super-large stellar flares, by solar standards.

\begin{figure}
\centerline{\includegraphics[width=1.0\textwidth]{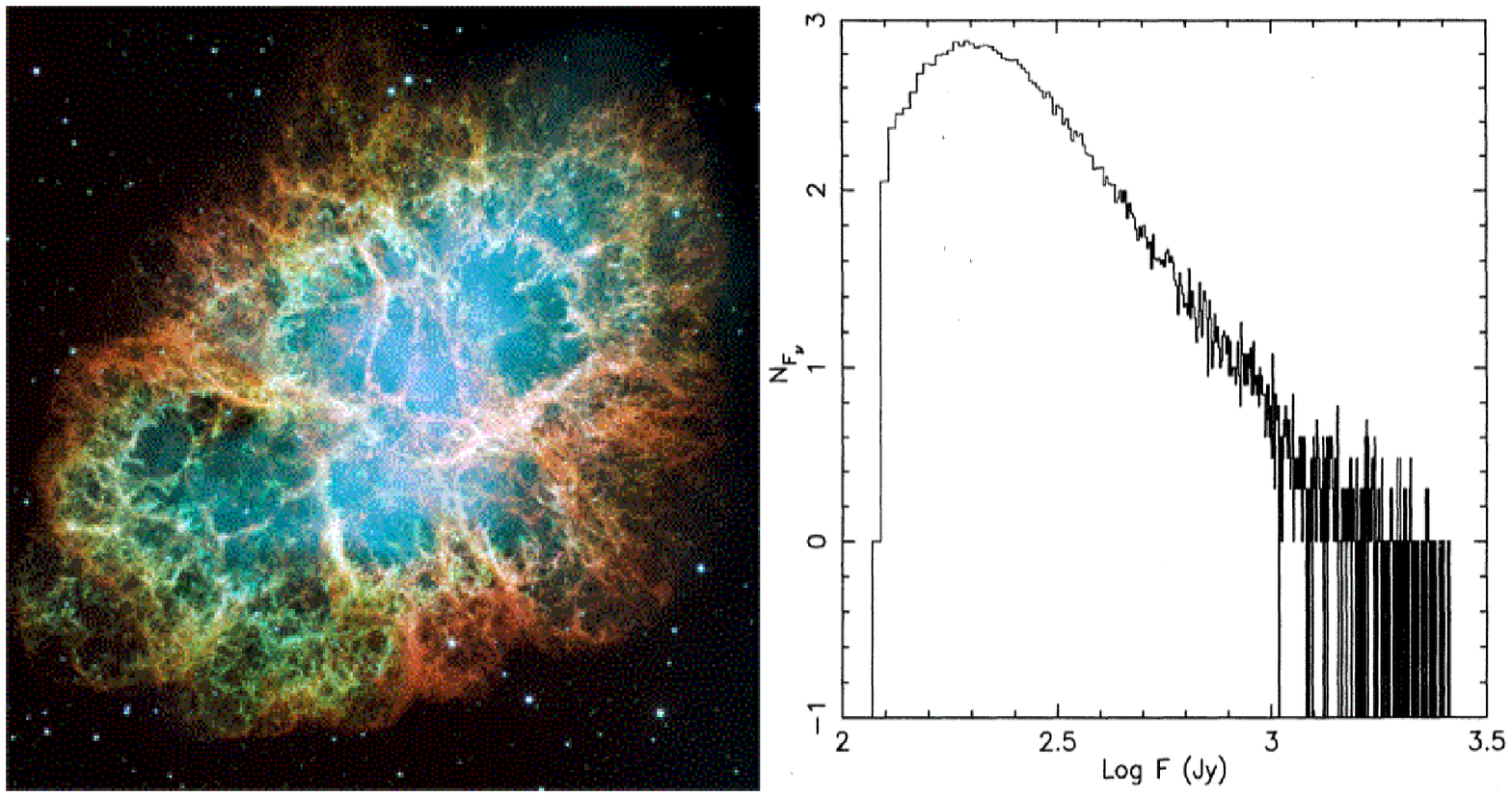}}
\captio{{\sl Left:} Crab nebula, which harbors the pulsar in the center
(photographed by Hubble Space Telescope, NASA).
{\sl Right:} Frequency distribution of giant-pulse flux densities measured
from the Crab pulsar, observed during 15-27 May 1991 with the Green Bank
43-m telescope at 1330, 800, and 812.5 MHz. The tail can be represented
by a powerlaw distribution $N_F \propto F^{-\alpha}$ with 
$\alpha=3.46\pm0.04$ for fluxes $F>200$ Jy (Lundgren et al.~1995).}
\end{figure}

\subsection{Pulsars}

Pulsars are fast-spinning neutron stars, which emit strictly periodic
signals in radio wavelengths, as well as occasional giant pulses that
represent glitches in the otherwise regular pulse amplitude and frequency.
The glitches in pulse amplitude and frequency shifts correspond to
large positive spin-ups of the neutron star, probably caused by sporadic
unpinning of vortices that transfer momentum to the crust. Conservation
of the angular momentum produces then an increase of the angular
rotation rate. Thus, these giant pulses reveal highly nonlinear energy
dissipation processes that can be considered as a SOC phenomenon and
we expect a powerlaw function for their size distribution.

Early measurements of the pulse height distribution of the Crab pulsar 
(NGC 0532 or PSR B0531+21) observed at 146 MHz were indeed found to have 
a powerlaw slope of $\alpha_P \approx \beta+1=3.5$ over a range of 
2.25 to 300 times the average pulse size, in a sample of 440 giant
pulses (Argyle and Gower 1972). Similar values were measured by
Lundgren et al.~(1995) in a sample of 30,000 giant pulses,  
with $\alpha_P \approx 3.06-3.36$ (Fig.~13.14, right).
While the Crab pulsar is the youngest known pulsar (born in the year 1054), 
PSR B1937+21 is an older pulsar with a 20 times faster period (1.56 ms) 
than the Crab pulsar (33 ms). Cognard et al. (1996) measured a powerlaw 
distribution with a slope of $\alpha_P \approx \beta+1=2.8\pm0.1$ 
from a sample of 60 giant pulses for this pulsar. 
Statistics of nine other pulsars revealed powerlaw slopes in
a much larger range of $\alpha_E=-0.13,...,2.4$ for the size distribution
of pulse glitches (Melatos et al.~2008), but those measurements were
obtained from much smaller samples of 6-30 giant pulses, and thus
represent small-number statistics.   

The typical value of $\alpha_P \approx 3.0$ found in two pulsars 
deviates significantly from the prediction ($\alpha_P = 1.67$) of 
our FD-SOC model, and thus requires either a different model or
more statistics from other pulsar cases. Preliminary values of
small samples from other pulsars indicate a wide range 
($\alpha_E=-0.13,...,2.4$, Melatos et al.~2008) that do not point
towards a particular value that can be explained with a single model.

\begin{figure}
\centerline{\includegraphics[width=1.0\textwidth]{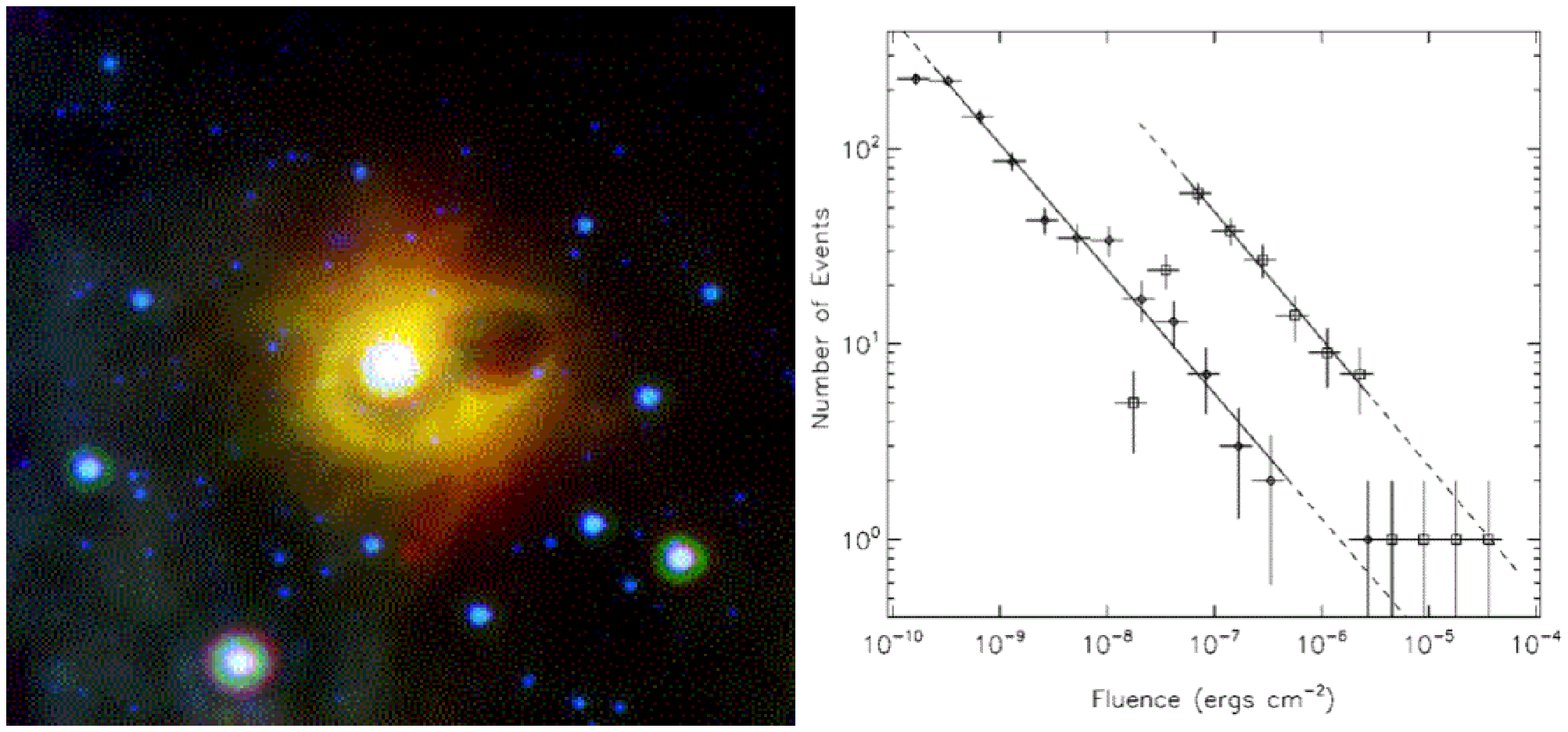}}
\captio{{\sl Left:} The Soft Gamma-Ray Repeater SGR 1900+14 in the
constellation of Aquila photographed with the Spitzer Space Telescope
in infrared (courtesy: NASA). {\sl Right:} Differential frequency 
distributions of the fluences of soft gamma-ray repeater SGR 1900+14 
observed with CGRO, RXTE, and ICE (Gogus et al.~1999).}
\end{figure}

\subsection{Soft Gamma-Ray Repeaters}

Gamma-ray bursts were observed from a variety of astrophysical 
objects, such as neutron stars or black holes, but usually only one
burst has been observed from each object. An exception is a class
of objects that show repetitive emission at low-energy gamma-rays
($>25$ keV), which were termed {\sl soft gamma-ray repeaters (GRS)}.
Observations with the {\sl Compton Gamma Ray Observatory (CGRO)}
revealed four such SGR sources up to 1999, three in our galaxy and
one in the Magellanic Cloud). At least three of these SGR objects
were associated with slowly rotating, extremely magnetized neutron
stars that are located in supernova remnants (Kouveliotou et al.~1998,
1999). It is believed that these soft gamma-ray bursts occur from
neutron star crust fractures driven by the stress of an evolving,
ultrastrong magnetic field in the order of $B \gapprox 10^{14}$ G.

Occurrence frequency distributions of the fluence of soft gamma-ray 
repeaters were obtained from four SGR sources: a database of 837 gamma-ray
bursts from SGR 1900+14 during the 1998-1999 active phase showed a
powerlaw slope of $\alpha_E=1.66$ over 4 orders of magnitude
(Fig.~13.15; Gogus et al.~1999); and a combined database from
SGR 1806-20, using 290 events detected with the {\sl Rossi X-Ray Timing
Explorer}, 111 events detected with {\sl CGRO/BATSE}, and 134 events
detected with the {\sl International Cometary Explorer (ICE)}, showing
power laws with slopes of $\alpha_E$=1.43, 1.76, and 1.67 (Gogus
et al.~2000). These measurements agree remarkably well with the
frequency distributions predicted by the FD-SOC model ($\alpha_E=1.50$)
as well as with those observed during solar flares, which were also
observed at the same hard X-ray energies of $>25$ keV. However, the
physical energy dissipation mechanism may be quite different in a
solar-like star and a highly magnetized neutron star,
given the huge difference in magnetic field strengths ($B \approx 10^2$ G
for solar flares versus $B \approx 10^{14}$ G in a magnetar), although
magnetic reconnection processes could be involved in both cases.
Nevertheless, soft gamma-ray repeaters have been interpreted as a
SOC system (Gogus et al.~1999), in terms of a neutron star crustquake
model (Thompson and Duncan 1996), in analogy to the SOC interpretation
of earthquakes.

\begin{figure}
\centerline{\includegraphics[width=1.0\textwidth]{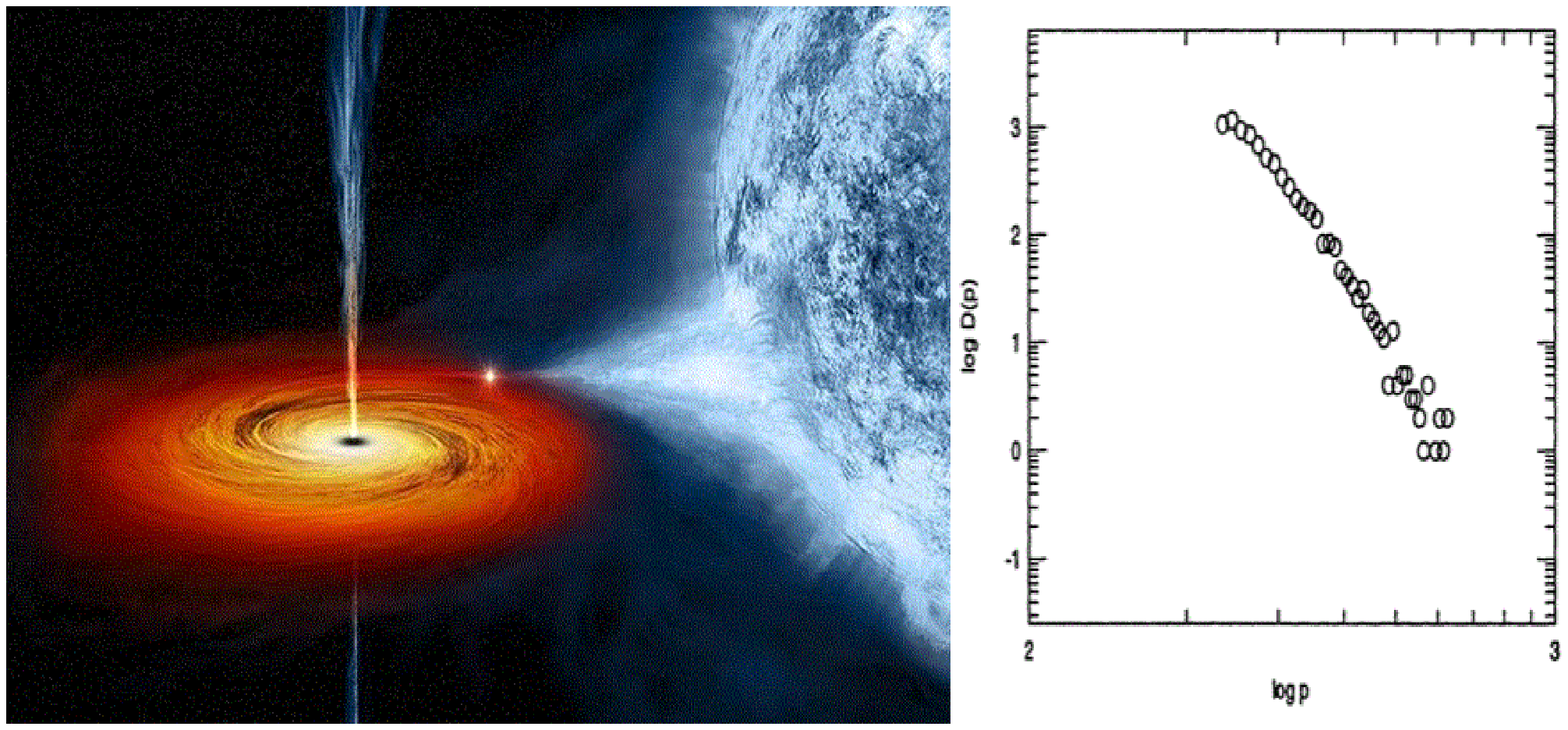}}
\captio{{\sl Left:} This artist concept of Cygnus X-1 shows the black hole
drawing material from a companion star (right) into a hot, swirling 
accretion disk that surrounds the (invisible) black hole
(Courtesy of Chandra X-Ray Observatory, NASA). {\sl Right:} Observed
frequency distribution of the peak intensities of pulses in the light 
curve of the black-hole object Cygnus X-1, exhibiting a powerlaw
slope of $\alpha_P \approx 7.1$ (Negoro et al.~1995; Mineshige and
Negoro 1999).}
\end{figure} 

\subsection{Black Hole Objects}

Cygnus X-1, a galactic X-ray source in the constellation Cygnus,
is the first X-ray source that has been widely accepted to be a black-hole
candidate. The mass of Cygnus X-1 is estimated to be about 14.8 solar
masses and it has been inferred that the object with (an event horizon at) 
a radius of 26 km is far too compact to be a normal star. Cygnus X-1 is
a high-mass X-ray binary star system, which draws mass from a blue
supergiant variable star (HDE 226868) in an orbit of 0.2 AU around
the black hole. The stellar wind of this blue companion star swirls
mass onto an accretion disk around the black hole (Fig.~13.16).
The X-ray time profile from Cygnus X-1 reveals time variability
down to 1 ms, which is attributed to X-ray pulses from matter 
infalling toward the black hole and the resulting turbulence in
the accretion disk. 

Observations of the X-ray light curve of Cygnus X-1 with Ginga
exhibit a complex power spectrum that entails at least 3 piece-wise
powerlaw sections, which have been interpreted as a superposition
of multiple 1/f-noise spectra (Takeuchi et al.~1995). The occurrence frequency
distribution of peak intensities shows a powerlaw-like function
with a steep slope of $\alpha_P \approx 7.1$ (Fig.~13.16 right;
Negoro et al.~1995; Mineshige and Negoro 1999). All these properties
have been modeled with a sophisticated cellular automaton model
in the framework of the SOC concept. Infalling mass lumps in the 
accretion disk are thought to trigger turbulent instabilities in
the neighborhood of an infall side, which propagate avalanche-like
and produce hard X-rays either by collisional bremsstrahlung or
some other magnetically driven instability (e.g., a magnetic
reconnection process). The cellular automaton simulations 
(Takeuchi et al.~1995, Mineshige and Negoro 1999) were able to reproduce
steep powerlaw slopes of the peak fluxes in the range of $\alpha_P
\approx 5.6-11.5$, depending on the effect of enhanced mass transfer
by gradual diffusion in addition to the avalanche-like shots, and this
way could reproduce the observations. 

We note that the observed
steep powerlaw slopes of peak fluxes ($\alpha_P \approx 7.1$) exceed
the predictions of the FD-SOC model ($\alpha_P=1.67$) by far.
Such a steep slope, $\alpha_P=1+2/(3\gamma) \approx 7$, can be produced 
by an extremely weak dependence of the X-ray flux $F$ on the avalanche
volume $V$, i.e., $F \propto V^\gamma$ with $\gamma \approx 1/9$, 
which is different from the flux-volume scaling law of optically-thin
soft X-ray or EUV emission ($\gamma \approx 1$) generally observed
in astrophysical sources, and thus may indicate a nonthermal emission 
mechanism. Alternatively, the very steep powerlaw slope of $\alpha_P$ 
could by part of an exponential frequency distribution near the upper
cutoff, which could only be proven by sampling peak fluxes with higher
sensitivity. Simultaneous modeling of the observed occurrence frequency
distributions of time scales, peak fluxes, and power spectra in terms
of a SOC model may reveal the underlying physical scaling law of the
emission mechanism in black-hole accretion disks.

\begin{figure}
\centerline{\includegraphics[width=1.0\textwidth]{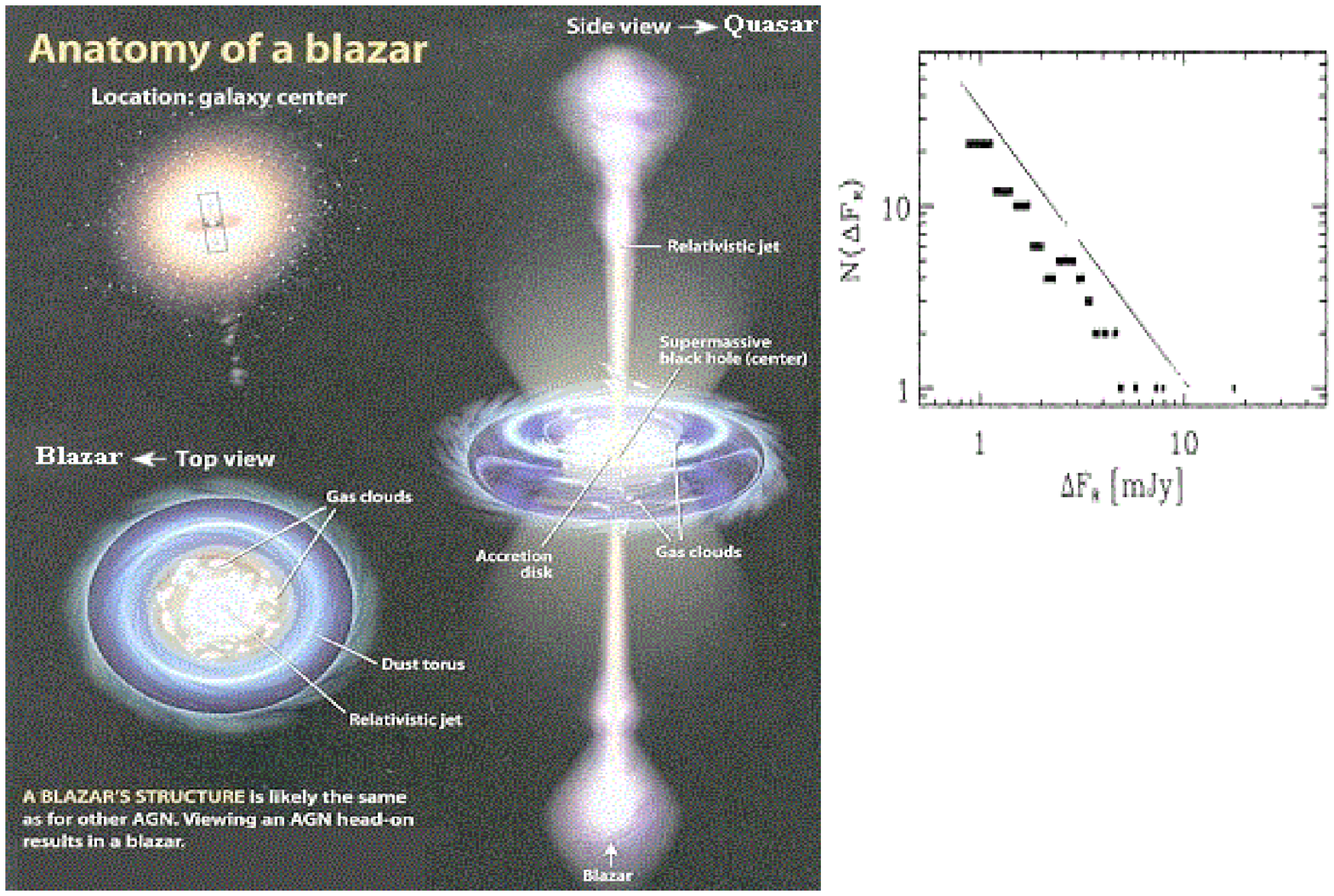}}
\captio{{\sl Left:} The anatomy of a blazar is shown in a top view
and side view, having the rotation axis with the relativistic
jet oriented toward Earth or orthogonal to the line-of-sight 
(Courtesy of The Encylopedia of Science).
{\sl Right:} Frequency distribution of peak fluxes of flaring events
in blazar GC 0109+224, including fluxes above a 3$\sigma$-threshold, 
fitted with a powerlaw function $N(P) \propto P^{-1.55}$ (Ciprini
et al.~2003).}
\end{figure}

\subsection{Blazars}

Blazars (blazing quasi-stellar objects) are a special subgroup of
quasars. This group includes BL Lacertae objects, high polarization 
quasars, and optically violent variables. They are believed to be 
active galactic nuclei whose jets are aligned within $\lapprox
10^\circ$ of our line-of-sight (Fig.~13.17 left). Because we are
observing ``down'' the jet direction we observe a large degree of 
variability and apparent superluminal speeds from the jet-aligned
emission. The structure of blazars, like all
active galactic nuclie, is thought to be powered by material falling
onto the supermassive black hole at the center of the host galaxy,
which emits highly intermittent gyrosynchrotron emission (in radio
wavelengths), inverse Compton emission (in X-rays and gamma-rays), 
and free-free bremsstrhalung emission (in soft X-rays), 
modulated by the variable rate of matter infalling into the
accretion disk of the central black hole. 

The optical variability of blazar GX 0109+224 was monitored and
the light curves were found to exhibit flickering and shot noise,
with a power spectrum $P(\nu) \propto \nu^{-p}$ with power
index of $p=1.57-2.05$ (Ciprini et al.~2003). The occurrence
frequency distribution of peak fluxes of flare events was found
to have a powerlaw slope of $\alpha_p=1.55$ (Fig.~13.17, right;
Ciprini et al.~2003), which is close to the prediction of the FD-SOC 
model ($\alpha_p =1.67$). Thus, the highly-variable blazar emission 
was interpreted in terms of SOC models. The fact that the peak
size distribution of radio emission observed in the blazar agrees 
with the prediction of the FD-SOC model is consistent with 
a near-proportional radio flux-volume scaling, i.e., $F \propto
V^\gamma$ with $\gamma \approx 1$, which is generally the case
for gyro-synchrotron emission. This is different from the flux
scaling of emission observed from the black hole Cygnus X-1. 
Thus, SOC statistics allows us to discriminate between different 
physical emission mechanisms in black holes and blazars. 

\begin{figure}
\centerline{\includegraphics[width=0.47\textwidth]{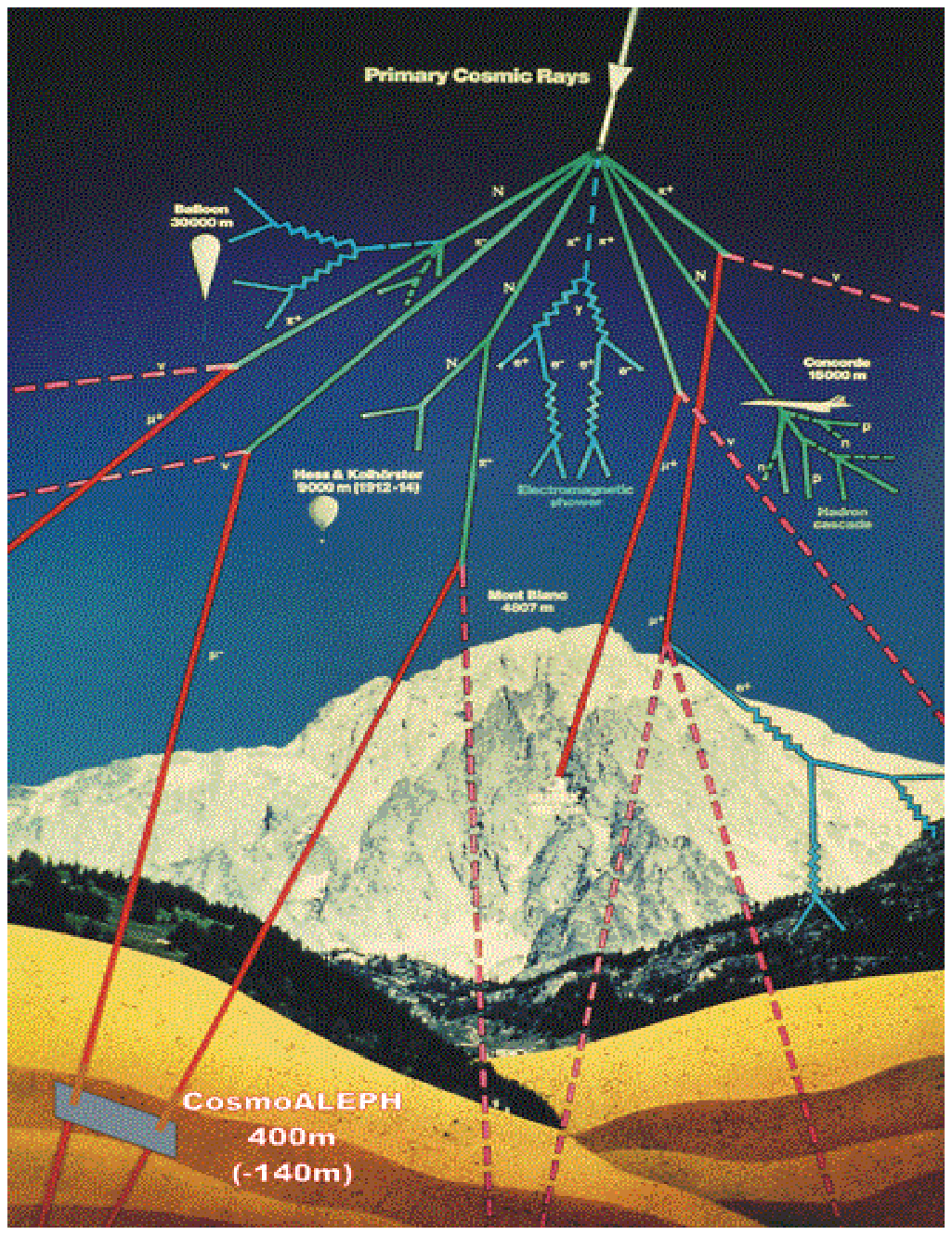},
            \includegraphics[width=0.53\textwidth]{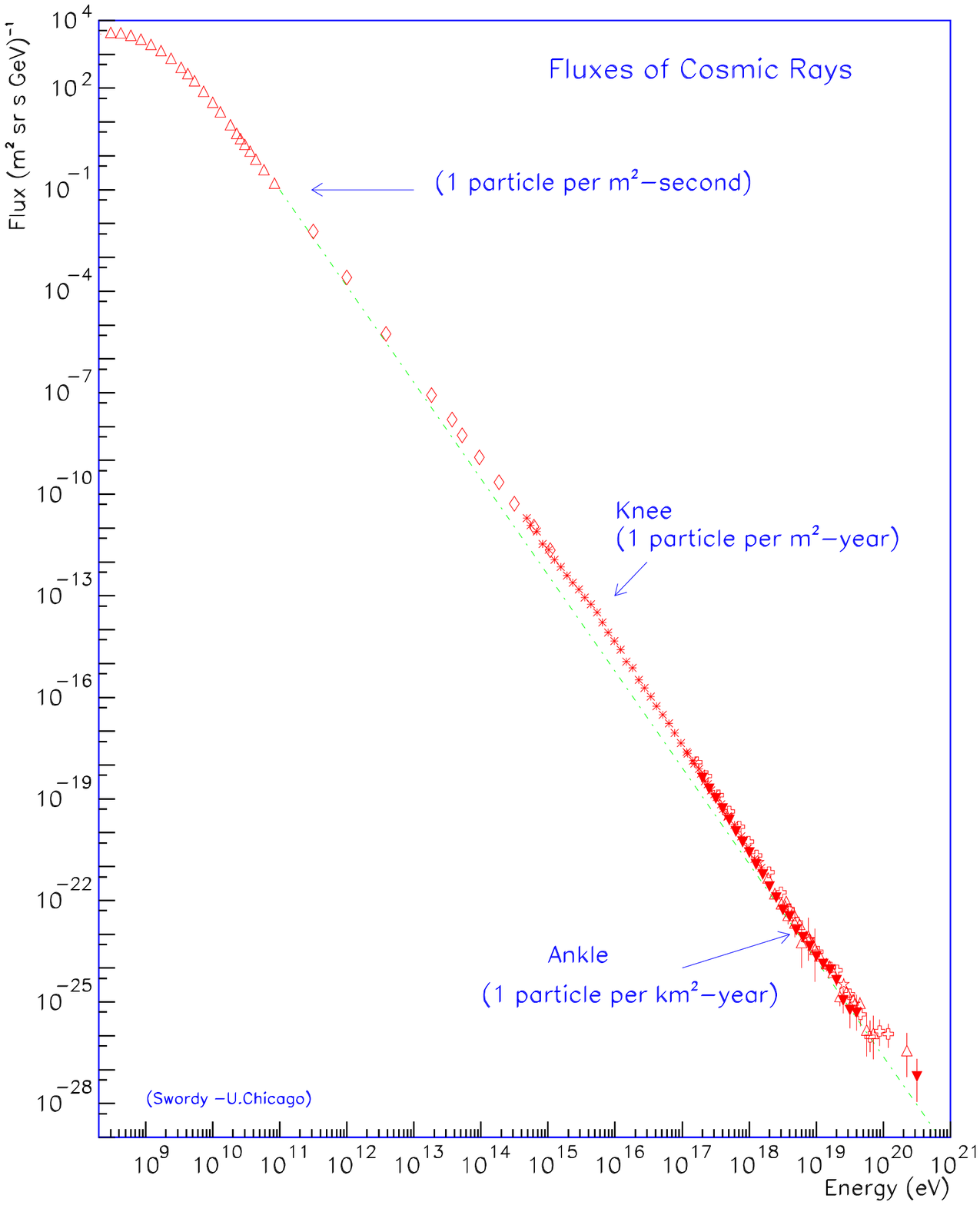}}
\captio{{\sl Left:} The diagram shows the cascade shower of a primary
cosmic ray particle, which produces in the Earth's atmosphere a shower of
secondary particles, which can be detected by an unterground detector,
such as CosmoALEPH (courtesy of CosmoALEPH Team).
{\sl Right:} Cosmic ray spectrum in the energy range of $E=10^9-10^{21}$ eV,
covering over 12 orders of magnitude. There is a ``knee'' in the
spectrum around $E \approx 10^{16}$ eV, which separates
cosmic rays originating within our galaxy (at lower energies) and
those from outside the galaxy (at higher energies).
(Credit: Simon Swordy, University of Chicago).}
\end{figure}

\subsection{Cosmic Rays}

Cosmic rays are high-energy particles that have been accelerated
during a long journey through a large part of our universe, inside
our Milky way as well as from outside of our galexy. Their boosting
to the highest energies of up to $\lapprox 10^{21}$ eV can only 
occur by multi-step acceleration processes throughout the universe
over long time spans. The energy spectrum of cosmic rays, as it can
be measured from showers of secondary particles produced in the
Earth atmosphere (Fig.~13.18 left), exhibits a powerlaw-like energy
spectrum extending from $\approx 10^9$ eV (=1 GeV) over 12 orders
of magnitude up to $\lapprox 10^{21}$ erg (Fig.~13.18 right).
The average slope is $\alpha_E \approx 2.7$. However, a more
accurate model is a broken powerlaw with a ``knee'' in the spectrum
around $E \approx 10^{16}$ eV. The widely accepted interpretation
of this knee is that it separates the origin of cosmic rays from
inside and outside of our galaxy. The powerlaw slope above the
``knee'' steepens to $\alpha_E \approx 3.0-3.3$. Interestingly,
our FD-SOC model applied to the kinetic energy gain in a coherent
direct current (DC) electric field implies a proportionality of
$E_L \propto L$ (Eq.~13.28) for sub-relativistic energies, and thus 
predicts an energy distribution of $N(E_L) \propto E_L^{-3}$, 
similar to the cosmic ray spectrum. Of course, cosmic rays are
highly relativistic and are likely to be produced by many $(n)$
acceleration phases. However, even if each acceleration phase
is local and has a relatively small length of $l \ll L$, the
energy gain of the particle would add up linearly with increasing
travel time and travel distance $\sum l$ and could still fulfill 
the proportionality,
$$
	E_L \propto \sum_{i=1}^n E_l \propto \sum_{i=1}^n l \propto L \ ,
	\eqno(13.34)
$$
and end up with an energy spectrum of $N(E_L) \propto L^{-3}$. 

So, can we understand the acceleration of a cosmic ray as a SOC process?
The lattice grid would cover a large fraction of the extragalactic
space of our universe, the ensemble of cosmic-ray particles would
represent an avalanche that nonlinearly dissipates energy from
local acceleration processes (such as by the first-order Fermi 
acceleration process between intergalactic or interstellar magnetic
clouds), which are self-organizing in the sense that accelerating
fields (of magnetic clouds) are constantly restored by the galactic
and interstellar dynamics. 

Our fractal-diffusive spatio-temporal
relationship ($r(t) \propto t^{\beta/2}$, Eq.~13.8) further predicts
a random-walk through the universe and an age $T$ that scales as 
$L \propto T^{1/2}$ with the straight travel distance $L$ from 
the point of origin of the cosmic ray particle. Using the ``knee''
in the cosmic-ray energy spectrum (at $E_{gal} \approx 10^{16}$ eV)
as a calibration for the distance of the Earth to the center of
our galaxy ($L_{gal} \approx 50$ light years $\approx 5 \times 10^{22}$ cm),
we can estimate the straight length scale over which the cosmic ray
particle travelled by random walk
$$
	L_{max} \approx L_{gal} \left( {E_{max} \over E_{gal}} \right) \ ,
	\eqno(13.35)
$$
which yields $L_{max} \approx 5 \times 10^{27}$ erg, which corresponds
to about 10\% of the size of our universe $r_{uni} \approx 4 \times 10^{28}$. 
Since the acceleration efficiency is different in galactic and extragalactic
space, the diffusion coefficient of the random walk is also different and
therefore we expect a different powerlaw slope in the energy spectrum
produced in these two regimes. 

\section{Conclusions}

In this chapter we generalized the fractal-diffusive self-organized criticality
(FD-SOC) model in terms of four fundamental parameters: (i) the Euclidean
dimension $S$, (ii) the fractal dimension $D_S$ of the spatial SOC avalanche 
structure, (iii) the diffusion index $\beta$ that includes both sub-diffusion
and super-diffusion, and (iv) the energy-volume scaling law with powerlaw
index $\gamma$. This model predicts powerlaw functions for the occurrence
frequency distributions of the SOC model, and moreover predicts their powerlaw
slope as a function of the four fundamental parameters. For a Euclidean
dimension of $S=3$, a mean fractal dimension of $D_S \approx (1+S)/2=2$,
classical diffusion ($\beta=1$), and linear flux-volume scaling ($\gamma=1$),
our generalized FD-SOC model predicts then the following powerlaw slopes:
$\alpha_L=3$ for length scales, $\alpha_A=2.333$ for areas, $\alpha_T=2$
for durations, $\alpha=1.667$ for peak fluxes, and $\alpha_E=1.5$ for 
fluences or total energies of the SOC avalanches.

Comparing these theoretical predictions with the observed powerlaws of size
distributions in astrophysical systems (summarized in Table 13.2) we find
acceptable agreement for the cases of lunar craters, asteroid belts, Saturn
rings, outer radiation belt electron bursts, solar flares, soft gamma-ray
repeaters, and blazars, if we apply the linear flux-volume scaling. 
Discrepancies are found for magnetospheric substorms, stellar flares, 
pulsar glitches, black holes, and cosmic rays, which apparently require
a nonlinear flux-volume scaling. Pulsar glitches and cosmic rays can
indeed be modeled by assuming a linear energy-length scaling, which leads
to energy spectra of $N(E) \approx E^{-3}$. Black-hole pulses have very
steep size spectra, which indicates a quenching or saturation process that
prevents a large variation of pulse amplitudes. Magnetospheric substorms
and solar energetic particles have the flattest size distributions, which
possibly can be explained by a selection effect with a bias for the
largest events. In conclusion, the generalized FD-SOC model can explain
a large number of astrophysical observations and can discriminate between
different scaling laws of astrophysical observables. We envision that
more refined scaling laws between astrophysical observables will be
developed that are consistent with the observed size distributions, and
this way will provide the ultimate predictive power for SOC models.

\begin{table}[t]
\begin{center}
\normalsize
\captio{Summary of theoretically predicted and observed powerlaw
indices of size distributions in astrophysical systems.}
\medskip
\begin{tabular}{|l|l|l|l|l|l|}
\hline
		                 & Length     & Area        & Duration    & Peak flux   & Fluence    \\
                                 & $\alpha_L$ & $\alpha_A$,
                                                $\alpha_{th,A}$ & $\alpha_T$  & $\alpha_P$  & $\alpha_E$ \\
\hline
FD-SOC Theory		         & {\bf 3.0}  & {\bf 2.33}  & {\bf 2.0}   & {\bf 1.67}  & {\bf 1.50} \\
\hline
\underbar{Lunar craters:}        &            &             &             &             &            \\
Mare Tranquillitatis $^1)$       & 3.0        &             &             &             &            \\
Meteorites and debris $^2)$      & 2.75       &             &             &             &            \\
\hline
\underbar{Asteroid belt:}        &            &             &             &             &            \\
{\sl Spacewatch Surveys}$^3)$    & 2.8        &             &             &             &            \\
{\sl Sloan Survey}$^4)$          & 2.3-4.0    &             &             &             &            \\
{\sl Subaru Survey}$^5)$         & 2.3        &             &             &             &            \\
\hline
\underbar{Saturn ring:}          &            &             &             &             &            \\
Voyager 1$^6)$                   & 2.74-3.11  &             &             &             &            \\
\hline
\underbar{Magnetosphere:}        &            &             &             &             &            \\                 
Active substorms$^7)$            &            &$1.21\pm0.08$&             &$1.05\pm0.08$&           \\
Quiet substorms$^7)$             &            &$1.16\pm0.03$&             &$1.00\pm0.02$&           \\
Substorm flow bursts$^8)$        &            &             &$1.59\pm0.07$&             &            \\
AE index bursts$^9)$             &            &             & 1.24        &             &            \\
AU index bursts$^{10})$          &            &             & 1.3         &             &            \\
Outer radiation belt$^{11})$     &            &             &             & 1.5-2.1     &            \\
\hline
\underbar{Solar Flares:}         &            &             &             &             &            \\                 
ISEE-3, HXR$^{12}$               &            &             & 1.88-2.73   & 1.75-1.86   & 1.51-1.62  \\
HXRBS/SMM, HXR$^{13}$            &            &             &$2.17\pm0.05$&$1.73\pm0.01$&$1.53\pm0.02$\\          
BATSE/CGRO, HXR$^{14}$           &            &             & 2.20-2.42   & 1.67-1.69   & 1.56-1.58  \\          
RHESSI, HXR$^{15}$               &            &             & 1.8-2.2     & 1.58-1.77   & 1.65-1.77  \\          
Yohkoh, SXR$^{16}$               & 1.96-2.41  & 1.77-1.94   &             & 1.64-1.89   & 1.4-1.6    \\
GOES, SXR$^{17}$                 &            &             & 2.0-5.0     & 1.86-1.98   & 1.88       \\
SOHO/EIT, EUV$^{18}$             &            & 2.3-2.6     & 1.4-2.0     &             &            \\
TRACE, EUV$^{19}$                & 2.50-2.75  & 2.4-2.6     &             & 1.52-2.35   & 1.41-2.06  \\
AIA/SDO, 335 A, EUV$^{20}$       & 1.96       &             & 2.17        & 1.34        &            \\
Microwave bursts$^{21}$          &            &             &             & 1.2-2.5     &            \\                 
Type III bursts$^{22}$           &            &             &             & 1.26-1.91   &            \\                 
Solar energetic particles$^{23}$ &            &             &             & 1.10-2.42   & 1.27-1.32  \\                 
\hline
\underbar{Stellar Flares:}       &            &             &             &             &            \\                 
Flare stars (reported)$^{24}$    &            &             &             &             &$2.17\pm0.25$\\
Flare stars (powerlaw fit)$^{24}$&            &             &             &             &$1.85\pm0.31$\\
Flare stars (cumulative fit)$^{24}$&          &             &             &             &$1.93\pm0.35$\\
\hline
\underbar{Astrophysical Objects:}&            &             &             &             &            \\                 
Crab pulsar$^{25}$               &            &             &             & 3.06-3.50   &            \\
PSR B1937+21$^{26}$              &            &             &             & $2.8\pm0.1$ &            \\
Soft Gamma-Ray repeaters$^{27}$  &            &             &             &             &$1.43-1.76$ \\
Cygnus X-1 black hole$^{28}$	 &            &             &             & 7.1         &            \\
Blazar GC 0109+224$^{29}$        &            &             &             & 1.55        &            \\
Cosmic rays$^{30}$		 &            &             &             &             &$2.7-3.3$   \\
\hline
\end{tabular}
\end{center}
\end{table}

\clearpage
{\footnotesize \underbar{References to Table 13.2:}
$^1)$ Cross (1966); 
$^2)$ Sornette (2004); 
$^3)$ Jedicke and Metcalfe (1998);
$^4)$ Ivezic et al.~(2001); 
$^5)$ Yoshida et al.~(2003), Yoshida and Nakamura (2007);
$^6)$ Zebker et al.~(1985), French and Nicholson (2000);
$^7)$ Lui et al.~(2000); 
$^8)$ Angelopoulos et al.~(1999);
$^9)$ Takalo (1993), Takalo et al.~(1999);
$^{10})$ Freeman et al.~(2000); Chapman and Watkins (2001);
$^{11})$ Crosby et al.~(2005)
$^{12})$ Lu et al.~(1993), Lee et al.~(1993);
$^{13})$ Crosby et al.~(1993);
$^{14})$ Aschwanden ~(2011a,b);
$^{15})$ Christe et al.~(2008), Lin et al.~(2001), Aschwanden ~(2011a,b);
$^{16})$ Shimizu (1995), Aschwanden and Parnell~(2002);
$^{17})$ Lee et al.~(1995), Feldman et al.~(1997), Veronig et al.~(2002a,b), 
	 Aschwanden and Freeland (2012);
$^{18})$ Krucker and Benz (1998), McIntosh and Gurman (2005);
$^{19})$ Parnell and Jupp (2000), Aschwanden et al. 2000, Benz and Krucker (2002), 
         Aschwanden and Parnell (2002), Georgoulis et al.~(2002); 
$^{20})$ Aschwanden (2012b)
$^{21})$ Akabane (1956), Kundu (1965), Kakinuma et al.~(1969), Das et al.~(1997), Nita et al.~(2002);
$^{22})$ Fitzenreiter et al.~(1976), Aschwanden et al.~(1995), Das et al.~(1997), Nita et al.~(2002);
$^{23})$ Van Hollebeke et al.~(1975), Belovsky and Ochelkov (1979), Cliver et al. (1991),
	 Gabriel and Feynman (1996), Smart and Shea (1997), Mendoza et al. (1997), 
	 Miroshnichenko et al.~(2001), Gerontidou et al.~(2002); 
$^{23})$ Gabriel and Feynman (1996);
$^{24})$ Robinson et al.~(1999), Audard et al.~(2000), Kashyap et al.~(2002),
	 G\"udel et al.~(2003), Arzner and G\"udel (2004), Arzner et al.~(2007), Stelzer et al.~(2007); 
$^{25})$ Argyle and Gower (1972), Lundgren et al.~(1995);
$^{26})$ Cognard et al.~(1996);
$^{27})$ Gogus et al.~(1999, 2000);
$^{28})$ Negoro et al.~(1995), Mineshige and Negoro (1999);
$^{29})$ Ciprini et al.~(2003);
$^{30})$ e.g., Fig.~13.18 (courtesy of Simon Swordy, Univ.Chicago).}

\section{References}

\ref{Akabane, K. 1956,
        {\sl Some features of solar radio bursts at around 3000 Mc/s},
        Publ. Astron. Soc. Japan {\bf 8}, 173-181.}
\ref{Angelopoulos, V., Mukai, T., and Kokubun, S. 1999,
        {\sl Evidence for intermittency in Earth's plasma sheet and
        implications for self-organized criticality},
        Phys. Plasmas {\bf 6/11}, 4161-4168.}
\ref{Arzner, K. and G\"udel, M. 2004,
        {\sl Are coronae of magnetically active stars heated by flares?
        III. Analytical distribution of superposed flares},
        Astrophys. J. {\bf 602}, 363-376.}
\ref{Arzner, K., G\"udel, M., Briggs, K., Telleschi, A., and Audard, M. 2007,
        {\sl Statistics of superimposed flares in the Taurus molecular cloud},
        Astron. Astrophys. {\bf 468}, 477-484.}
\ref{Aschwanden, M.J., Benz, A.O., Dennis, B.R., and Schwartz, R.A. 1995,
        {\sl Solar electron beams detected in hard X-rays and radio waves},
        Astrophys. J. {\bf 455}, 347-365.}
\ref{Aschwanden, M.J., Tarbell, T., Nightingale, R., Schrijver, C.J.,
        Title, A., Kankelborg, C.C., Martens, P.C.H., and Warren, H.P.
        2000, {\sl Time variability of the quiet Sun observed with TRACE:
        II. Physical parameters, temperature evolution, and energetics
        of EUV nanoflares}, Astrophys. J. {\bf 535}, 1047-1065.}
\ref{Aschwanden, M.J. and Parnell, C.E. 2002,
        {\sl Nanoflare statistics from first principles: fractal geometry
        and temperature synthesis}, Astrophys. J. {\bf 572}, 1048-1071.}
\ref{Aschwanden, M.J. 2011a,
 	{\sl Self-Organized Criticality in Astrophysics. The Statistics 
	of Nonlinear Processes in the Universe}, ISBN 978-3-642-15000-5, 
	Springer-Praxis: New York, 416p.}
\ref{Aschwanden, M.J. 2011b,
	{\sl The state of self-organized criticality of the Sun during
	the last three solar cycles. I. Observations},
	Solar Phys 274, 99-117.}
\ref{Aschwanden,M.J. 2012a,
 	{\sl A statistical fractal-diffusive avalanche model of a 
	slowly-driven self-organized criticality system},
 	Astron.Astrophys. 539, A2 (15 p).}
\ref{Aschwanden,M.J. 2012b, 
 	{\sl The spatio-temporal evolution of solar flares observed with 
	AIA/SDO: Fractal diffusion, sub-diffusion, or logistic growth ?}
	ApJ (subm.).}
\ref{Aschwanden,M.J. and Freeland,S.L. 2012, 
	{\sl Automated solar flare statistics in soft X-rays over 37 years 
	of GOES observations - The invariance of self-organized criticality
	during three solar cycles}, ApJ (in press).}
\ref{Audard, M., G\"udel, M., Drake, J.J., and Kashyap, V.L. 2000,
        {\sl Extreme-ultraviolet flare activity in late-type stars}
        Astrophys. J. {\bf 541}, 396-409.}
\ref{Bai, T. 1993,
        {\sl Variability of the occurrence frequency of solar flares as a
        function of peak hard X-ray rate},
        Astrophys. J. {\bf 404}, 805-809.}
\ref{Belovsky, M.N., and Ochelkov, Yu. P. 1979,
        {\sl Some features of solar-flare electromagnetic and corpuscular
        radiation production}, Izvestiya AN SSR, Phys. Ser. {\bf 43}, 749-752.}
\ref{Benz, A.O. and Krucker, S. 2002,
        {\sl Energy distribution of microevents in the quiet solar corona},
        Astrophys. J. {\bf 568}, 413-421.}
\ref{Biesecker, D.A. 1994,
        {\sl On the occurrence of solar flares observed with the
        Burst and Transient Source Experiment (BATSE)},
        PhD Thesis, University of New Hampshire.}
\ref{Chapman, S.C. and Watkins, N. 2001,
        {\sl Avalanching and self-organised criticality, a paradigm
        for geomagnetic activity?},
        Space Sci. Rev. {\bf 95}, 293-307.}
\ref{Christe, S., Hannah, I.G., Krucker, S., McTiernan, J., and Lin, R.P.
        2008, {\sl RHESSI microflare statistics. I. Flare-finding and
        frequency distributions}, Astrophys. J. {\bf 677}, 1385-1394.}
\ref{Ciprini, S., Fiorucci, M., Tosti, G., and Marchili, N. 2003,
        {\sl The optical variability of the blazar GV 0109+224. Hints of
        self-organized criticality},
        in {\sl High energy blazar astronomy}, ASP Conf. Proc. {\bf 229},
        (eds. L.O. Takalo and E. Valtaoja), ASP: San Francisco, p.265.}
\ref{Cliver, E., Reames, D., Kahler, S., and Cane, H. 1991,
        {\sl Size distribution of solar energetic particle events},
        Internat. Cosmic Ray Conf. 22nd, Dublin, LEAC A92-36806 15-93,
        NASA:Greenbelt, p. 2:1-4.}
\ref{Crosby, N.B., Aschwanden, M.J., and Dennis, B.R. 1993,
        {\sl Frequency distributions and correlations of solar X-ray
        flare parameters}, Solar Phys. {\bf 143}, 275-299.}
\ref{Crosby, N.B., Meredith, N.P., Coates, A.J., and Iles, R.H.A. 2005,
        {\sl Modelling the outer radiation belt as a complex system in a
        self-organised critical state},
        Nonlinear Processes in Geophysics {\bf 12}, 993-1001.}
\ref{Cross, C.A. 1966,
        {\sl The size distribution of lunar craters},
        MNRAS {\bf 134}, 245-252.}
\ref{Das, T.K., Tarafdar, G., and Sen, A.K. 1997,
        {\sl Validity of power law for the distribution of intensity
        of radio bursts}, Solar Phys. {\bf 176}, 181-184.}
\ref{Feldman, U., Doschek, G.A., and Klimchuk, J.A. 1997,
        {\sl The occurrence rate of soft X-ray flares as a function of
        solar activity}, Astrophys. J. {\bf 474}, 511-517.}
\ref{Fitzenreiter, R.J., Fainberg, J., and Bundy, R.B. 1976,
        {\sl Directivity of low frequency solar type III radio bursts},
        Solar Phys. {\bf 46}, 465-473.}
\ref{Freeman, M.P., Watkins, N.W., and Riley, D.J. 2000,
        {\sl Evidence for a solar wind origin of the power law burst lifetime
        distribution of the AE indices}, Geophys. Res. Lett. {\bf 27}, 1087-1090.}
\ref{French, R.G. and Nicholson, P.D. 2000,
        {\sl Saturn's rings. II. Partice sizes inferred from stellar
        occultation data}, Icarus {\bf 145}, 502-523.}
\ref{Gabriel, S.B. and Feynman, J. 1996,
        {\sl Power-law distribution for solar energetic proton events},
        Solar Phys. {\bf 165}, 337-346.}
\ref{Georgoulis, M.K., Kluivin,R. and Vlahos,L. 1995,
        {\sl Extended instability criteria in iso\-tro\-pic and anisotropic
        energy avalanches}, Physica A {\bf 218}, 191-213.}
\ref{Gerontidou, M., Vassilaki, A., Mavromichalaki, H., and Kurt, V. 2002,
        {\sl Frequency distributions of solar proton events},
        J. Atmos. Solar-Terr. Physics {\bf 64/5-6}, 489-496.}
\ref{Gogus, E., Woods, P.M., Kouveliotou, C., van Paradijs, J.,
        Briggs, M.S., Duncan, R.C., and Thompson, C. 1999,
        {\sl Statistical properties of SGR 1900+14 bursts},
        Astrophys. J. {\bf 526}, L93-L96.}
\ref{Gogus, E., Woods, P.M., Kouveliotou, C., and van Paradijs, J. 2000,
        {\sl Statistical properties of SGR 1806-20 bursts},
        Astrophys. J. {\bf 532}, L121-L124.}
\ref{Greenberg,R., Davies, D.R., Harmann, W.K., and Chapman, C.R.,
	1977, Icarus 30, 769-779.}
\ref{G\"udel, M., Audard, M., Kashyap, V.L., and Guinan, E.F. 2003,
        {\sl Are coronae of magnetically active stars heated by flares?
        II. Extreme Ultraviolet and X-ray flare statistics and the
        differential emission measure distribution},
        Astrophys. J. {\bf 582}, 423-442.}
\ref{Hudson, H.S. 1991,
        {\sl Solar flares, microflares, nanoflares, and coronal heating},
        Solar Phys. {\bf 133}, 357-369.}
\ref{Ivezic, Z., Tabachnik, S., Rafikov, R., Lupton, R.H., Quinn, T.,
        Hammergren, M., Eyer, L., Chu, J., Armstrong, J.C., Fan, X.,
        Finlator, K., Geballe, T.R., Gunn, J.E., Hennessy, G.S., Knapp, G.R.,
        et al. (SDSS Collaboration) 2001,
        {\sl Solar system objects observed in the Sloan Digital Sky Survey
        Commissioning Data}, Astronomical J. {\bf 122}, 2749-2784.}
\ref{Jedicke, R. and Metcalfe, T.S. 1998,
        {\sl The orbital and absolute magnitude distributions of main belt
        asteroids}, Icarus {\bf 131/2}, 245-260.}
\ref{Kakinuma, T., Yamashita, T., and Enome, S. 1969,
        {\sl A statistical study of solar radio bursts a microwave
        frequencies}, Proc. Res. Inst. Atmos. Nagoya Univ. Japan, Vol.
        {\bf 16}, 127-141.}
\ref{Kashyap, V.L., Drake, J.J., G\"udel, M., and Audard, M. 2002,
        {\sl Flare heating in stellar coronae},
        Astrophys. J. {\bf 580}, 1118-1132.}
\ref{Kouveliotou,C., Dieters, S., Strohmayer, T., van Paradijs, J.,
        Fishman, G.J., Meegan, C.A., Hurley, K., Kommers, J.,
        Smith, I., Frail, D., Muakami, T. 1998,
        {\sl An X-ray pulsar with a superstrong magnetic field in the soft
        $\gamma$-ray repeater SGR 1806-20},
        Nature {\bf 393}, 235-237.}
\ref{Kouveliotou,C., Strohmayer, T., Hurley, K., van Paradijs, J.,
        Finger, M.H., Dieters, S., Woods, P., Thomson, C., and Duncan, R.C.
        1999, {\sl Discovery of a magnetar associated with the soft gamma
        ray repeater SGR 1900+14}, Astrophys. J. {\bf 510}, L115-L118.}
\ref{Krucker, S. and Benz, A.O. 1998,
        {\sl Energy distribution of heating processes in the quiet solar
        corona}, Astrophys. J. {\bf 501}, L213-L216.}
\ref{Kundu, M.R. 1965,
        {\sl Solar radio astronomy},
        Interscience Publication: New York, 660 p.}
\ref{Lee, T.T., Petrosian, V., and McTiernan, J.M. 1993,
        {\sl The distribution of flare parameters and implications
        for coronal heating}, Astrophys. J. {\bf 412}, 401-409.}
\ref{Lee, T.T., Petrosian, V., and McTiernan, J.M. 1995,
        {\sl The Neupert effect and the chromospheric evaporation model
        for solar flares}, Astrophys. J. {\bf 418}, 915-924.}
\ref{Lin, R.P., Feffer,P.T., and Schwartz,R.A. 2001,
        {\sl Solar Hard X-Ray Bursts and Electron Acceleration Down to 8 keV},
        Astrophys. J. {\bf 557}, L125-L128.}
\ref{Lu, E.T. and Hamilton, R.J. 1991,
        {\sl Avalanches and the distribution of solar flares},
        Astrophys. J. {\bf 380}, L89-L92.}
\ref{Lu, E.T., Hamilton, R.J., McTiernan, J.M., and Bromund, K.R. 1993,
        {\sl Solar flares and avalanches in driven dissipative systems},
        Astrophys. J. {\bf 412}, 841-852.}
\ref{Lui, A.T.Y., Chapman, S.C., Liou,K., Newell, P.T., Meng, C.I.,
        Brittnacher, M., and Parks, G.K. 2000,
        {\sl Is the dynamic magnetosphere an avalanching system?},
        Geophys. Res. Lett. {\bf 27/7}, 911-914.}
\ref{McIntosh, S.W. and Gurman, J.B. 2005,
        {\sl Nine years of EUV bright points},
        Solar Phys. {\bf 228}, 285-299.}
\ref{Mendoza, B., Melendez-Venancio, R., Miroshnichenko, L.I., and
        Perez-Enriquez, R. 1997,
        {\sl Frequency distributions of solar proton events},
        Proc. 25th Int. Cosmic Ray Conf. {\bf 1}, 81.}
\ref{Mineshige, S. and Negoro, H. 1999,
        {\sl Accretion disks in the context of self-organized criticality:
        How to produce 1/f fluctuations ?},
        in {\sl High energy processes in accreting black holes},
        ASP Conf. Ser. {\bf 161}, 113-128.}
\ref{Miroshnichenko, L.I., Mendoza, B., and Perez-Enriquez R. 2001,
        {\sl Size distributions of the $>$10 MeV solar proton events},
        Solar Phys. {\bf 202}, 151-171.}
\ref{Nishizuka, N., Asai, A., Takasaki, H., Kurokawa, H., and Shibata, K. 2009,
 	{\sl The Power-Law Distribution of Flare Kernels and Fractal Current 
	Sheets in a Solar Flare}, 
 	Astrophys. J. {\bf 694}, L74-L77.}
\ref{Negoro, H., Kitamoto, S., Takeuchi, M., and Mineshige, S. 1995,
        {\sl Statistics of X-ray fluctuations from Cygnus X-1:
        Reservoirs in the disk ?} Astrophys. J. {\bf 452}, L49-L52.}
\ref{Nita, G.M., Gary, D.E., Lanzerotti, L.J., and Thomson, D.J. 2002,
        {\sl The peak flux distribution of solar radio bursts},
        Astrophys. J. {\bf 570}, 423-438.}
\ref{Parnell,C.E. and Jupp,P.E. 2000,
        {\sl Statistical analysis of the energy distribution of nano\-flares
        in the quiet Sun}
        Astrophys. J. {\bf 529}, 554-569.}
\ref{Robinson, R.D., Carpenter, K.G., and Percival, J.W. 1999,
        {\sl A search for microflaring activity on dMe flare stars.
        II. Observations of YZ Canis Minoris},
        Astrophys. J. {\bf 516}, 916-923.}
\ref{Rosner, R., and Vaiana, G.S. 1978,
        {\sl Cosmic flare transients: constraints upon models for energy 
	storage and release derived from the event frequency distribution},
        Astrophys. J. {\bf 222}, 1104-1108.}
\ref{Shibata, K. and Yokoyama T. 1999,
        {\sl Origin of the universal correlation between the flare temperature
        and the emission measure for solar and stellar flares},
        Astrophys. J. {\bf 526}, L49-L52.}
\ref{Shibata, K. and Yokoyama T. 2002,
        {\sl A Hertzsprung-Russell-like diagram for solar/stellar flares and
        corona: emission measure versus temperature diagram},
        Astrophys. J. {\bf 577}, 422-432.}
\ref{Shimizu, T. 1995,
        {\sl Energetics and occurrence rate of active-region transient
        brightenings and implications for the heating of the active-region
        corona}, Publ. Astron. Soc. Japan {\bf 47}, 251-263.}
\ref{Smart, D.F. and Shea, M.A. 1997,
        {\sl Comment on the use of solar proton spectra in solar proton
        dose calculations}, in Proc. {\sl Solar-Terrestrial
        Prediction Workshop V}, Hiraiso Solar-Terrestrial Research Center,
        Japan, p.449.}
\ref{Sornette, D. 2004,
        {\sl Critical phenomena in natural sciences: chaos, fractals,
        self-organi\-za\-tion and disorder: concepts and tools},
        Springer, Heidelberg, 528 p.}
\ref{Stelzer, B., Flaccomio, E., Briggs, K., Micela, G., Scelsi, L,
        Audard, M., Pillitteri, I., and G\"udel, M. 2007,
        {\sl A statistical analysis of X-ray variability in pre-main
        sequence objects of the Taurus molecular cloud},
        Astron. Astrophys. {\bf 468}, 463-475.}
\ref{Takalo, J. 1993,
        {\sl Correlation dimension of AE data},
        Ph. Lic. Thesis, {\sl Laboratory report 3}, Dept. Physics,
        University of Jyv\"askyl\"a.}
\ref{Takalo, J., Timonem, J., Klimas, A., Valdivia, J., and Vassiliadis, D.
        1999, {\sl Nonlinear energy dissipation in a cellular automaton
        magnetotail field model}
        Geophys. Res. Lett. {\bf 26/13}, 1813-1816.}
\ref{Takeuchi, M., Mineshige, S., and Negoro, H. 1995,
        {\sl X-ray fluctuations from black-hole objects and self organization
        of accretion disks},
        Publ. Astron. Soc. Japan {\bf 47}, 617-627.}
\ref{Thompson, C. and Duncan, R.C. 1996,
        {\sl The soft gamma repeaters as very strongly magnetized neutron
        stars. II. Quiescent neutrino, X-ray, and Alfv\'en wave emission}
        Astrophys. J. {\bf 473}, 322-342.}
\ref{VanHollebeke, M.A.I., Ma Sung L.S., and McDonald F.B. 1975,
        {\sl The variation of solar proton energy spectra and
        size distribution with heliolongitude},
        Solar Phys. {\bf 41}, 189-223.}
\ref{Veronig, A., Temmer, M.,Hanslmeier, A., Otruba, W., and Messerotti, M.
        2002a, {\sl Temporal aspects and frequency distributions of solar
        X-ray flares}, Astron. Astrophys. {\bf 382}, 1070-1080.}
\ref{Veronig, A., Temmer, M., and Hanslmeier, A. 2002b,
        {\sl Frequency distributions of solar flares},
        Hvar Observatory Bulletin {\bf 26/1}, 7-12.}
\ref{Yoshida, F., Nakamura, T., Watanab, J., Kinoshita, D.,
        and Yamamoto, N., 2003,
        {\sl Size and spatial distributions of sub-km main-belt asteroids},
        Publ. Astron. Soc. Japan {\bf 55}, 701-715.}
\ref{Yoshida, F. and Nakamura T. 2007,
        {\sl Subary Main Belt Asteroid Survey (SMBAS) - Size and color
        distributions of small main-belt asteroids},
        Planet. Space Science {\bf 55}, 113-1125.}
\ref{Zebker, H.A., Maroufm, E.A., and Tyler, G.L. 1985,
        {\sl Saturn's rings particle size distributions for a thin layer
        model}, Ikarus {\bf 64}, 531-548.}



\end{document}